\documentclass[fleqn,usenatbib]{mnras}

\usepackage[T1]{fontenc}
\DeclareRobustCommand{\VAN}[3]{#2}
\let\VANthebibliography\thebibliography
\def\thebibliography{\DeclareRobustCommand{\VAN}[3]{##3}\VANthebibliography}

\usepackage{graphicx}   
\usepackage{amsmath}    
\usepackage{amssymb}    
\usepackage[flushleft]{threeparttable}

\newcommand  \kms      {\ifmmode {\rm km\,s}^{-1} \else km\,s$^{-1}$\fi}
\newcommand  \cc       {\hbox{cm$^{-3}$}}
\newcommand  \cmii     {\hbox{cm$^{-2}$}}

\newcommand  \ergs     {\ifmmode {\rm ergs\,s}^{-1} \else ergs s$^{-1}$\fi}
\newcommand  \ergcms   {\ifmmode {\rm ergs\,cm}^{-2}\,{\rm s}^{-1}
                        \else ergs\,cm$^{-2}$\,s$^{-1}$\fi}
\newcommand  \ergcmsA  {\ifmmode{\rm ergs\,cm}^{-2}\,{\rm s}^{-1}\,{\rm\AA}^{-1}
                        \else ergs\,cm$^{-2}$\,s$^{-1}$\,\AA$^{-1}$\fi}
\newcommand  \ergcmsHz {\ifmmode{\rm ergs\,cm}^{-2}\,{\rm s}^{-1}\,{\rm Hz}^{-1}
                        \else ergs\,cm$^{-2}$\,s$^{-1}$\,Hz$^{-1}$\fi}
\newcommand  \phcms    {\ifmmode {\rm ph\,cm}^{-2}\,{\rm s}^{-1}
                        \else ,ph\,cm$^{-2}$\,s$^{-1}$\fi}
\newcommand  \phcmsA   {\ifmmode {\rm ph\,cm}^{-2}\,{\rm s}^{-1}\,{\rm\AA}^{-1}
                        \else ph\,cm$^{-2}$\,s$^{-1}$\,\AA$^{-1}$\fi}
\def  \apj {ApJ}

\def  \mnras {MNRAS}

\def \aap   {A\&A}
\def \araa {AnnRevAstAp}

\def\micron{\ifmmode \mu{\rm m} \else $\mu$m\fi}
\def\mic{\ifmmode \mu{\rm m} \else $\mu$m\fi}
\def\kms{\ifmmode {\rm km\,s}^{-1} \else km\,s$^{-1}$\fi}
\def\Hubble{\ifmmode {\rm km\,s}^{-1}\,{\rm Mpc}^{-1}
        \else km\,s$^{-1}$\,Mpc$^{-1}$\fi}
\def\ergsec{\ifmmode {\rm ergs\;s}^{-1} \else ergs s$^{-1}$\fi}
\def\ergscm{\ifmmode {\rm ergs\,s}^{-1}\,{\rm cm}^{-2}
          \else ergs\,s$^{-1}$\,cm$^{-2}$\fi}
\def\ergscmA{\ifmmode {\rm ergs\,s}^{-1}\,{\rm cm}^{-2}\,{\rm \AA}^{-1}
          \else ergs\,s$^{-1}$\,cm$^{-2}$\,\AA$^{-1}$\fi}
\def\ergscmHz{\ifmmode {\rm ergs\,s}^{-1}\,{\rm cm}^{-2}\,{\rm Hz}^{-1}
          \else ergs\,s$^{-1}$\,cm$^{-2}$\,Hz$^{-1}$\fi}
\def\deg{\ifmmode ^{\rm o} \else $^{\rm o}$\fi}
\def\cc{\ifmmode {\rm cm}^{-3} \else cm$^{-3}$\fi}
\def\Msun{\ifmmode M_{\odot} \else $M_{\odot}$\fi}
\def\msun{\ifmmode M_{\odot} \else $M_{\odot}$\fi}
\def\msunyr{\ifmmode M_{\odot}\,yr^{-1} \else $M_{\odot}\,yr^{-1}$\fi}
\def\Msyr{\ifmmode M_{\odot}\,yr^{-1} \else $M_{\odot}\,yr^{-1}$\fi}
\def\Lsun{\ifmmode L_{\odot} \else $L_{\odot}$\fi}
\def\lsun{\ifmmode L_{\odot} \else $L_{\odot}$\fi}
\def\mseed{\ifmmode M_{seed} \else $M_{seed}$\fi}
\def\Mseed{\ifmmode M_{seed} \else $M_{seed}$\fi}
\def\qo{\ifmmode q_{0} \else $q_{0}$\fi}
\def\Ho{\ifmmode H_{0} \else $H_{0}$\fi}
\def\ho{\ifmmode h_{0} \else $h_{0}$\fi}
\def\qo{\ifmmode q_{0} \else $q_{0}$\fi}
\def\ao{\ifmmode a_{0} \else $a_{0}$\fi}
\def\to{\ifmmode t_{0} \else $t_{0}$\fi}
\def\Halpha{\ifmmode {\rm H}\alpha \else H$\alpha$\fi}
\def\ha{\ifmmode {\rm H}\alpha \else H$\alpha$\fi}
\def\Ha{\ifmmode {\rm H}\alpha \else H$\alpha$\fi}
\def\Pa{\ifmmode {\rm P}\alpha \else P$\alpha$\fi}
\def\Brg{\ifmmode {\rm Br}\gamma \else Br$\gamma$\fi}
\def\Hbeta{\ifmmode {\rm H}\beta \else H$\beta$\fi}
\def\hb{\ifmmode {\rm H}\beta \else H$\beta$\fi}
\def\Hb{\ifmmode {\rm H}\beta \else H$\beta$\fi}
\def\hg{\ifmmode {\rm H}\gamma \else H$\gamma$\fi}
\def\Hg{\ifmmode {\rm H}\gamma \else H$\gamma$\fi}
\def\Hgamma{\ifmmode {\rm H}\gamma \else H$\gamma$\fi}
\def\Hdelta{\ifmmode {\rm H}\delta \else H$\delta$\fi}
\def\Lya{\ifmmode {\rm Ly}\alpha \else Ly$\alpha$\fi}
\def\lya{\ifmmode {\rm Ly}\alpha \else Ly$\alpha$\fi}
\def\Lyb{\ifmmode {\rm Ly}\beta \else Ly$\beta$\fi}
\def\lyb{\ifmmode {\rm Ly}\beta \else Ly$\beta$\fi}
\def\hi{\ifmmode \mbox{{\rm H}\,{\sc i}} \else H\,{\sc i}\fi}

\def\heii{He{\sc ii}}

\def\ciii{\ifmmode {\rm C}{\sc iii}$]$ \else C{\sc iii}$]$\fi}
\def\civ{C{\sc iv}\,$\lambda1549$}

\def\o5007{[O\,{\sc iii}]\,$\lambda5007$}
\def  \Cloudy      {{\it Cloudy}}

\def \Lop      {${L_{5100}}$}
\def \Lopn      {${L_{5100,44}}$}
\def \Lopna     {${L_{5100,44}^{1/2}}$}

\def \LAGN{${L_{AGN}}$}
\def \LX{${L_{X}}$}
\def \Lx{${L_{X}}$}
\def \Ldisk{${L_{disk}}$}

\def \Ledd{$L/L_{Edd}$}
\def \D4000{${\rm D_n4000}$}
\def \d4000{${\rm D_n4000}$}

\def  \d4000       {D_{n}4000}
\def  \kms         {\hbox{km s$^{-1}$}}          
\def  \ergs        {\hbox{erg s$^{-1}$}}              
\def  \cc          {\hbox{cm$^{-3}$}}

\def  \cmii        {\hbox{cm$^{-2}$}}
\def  \mic         {$\mu$m}
\def  \La          {\ifmmode {\rm Ly}\alpha \else Ly$\alpha$\fi}
\def  \la          {\ifmmode {\rm Ly}\alpha \else Ly$\alpha$\fi}
\def  \Ka          {\ifmmode {\rm K}\alpha \else K$\alpha$\fi}
\def  \ka          {\ifmmode {\rm K}\alpha \else K$\alpha$\fi}
\def  \Lb          {\ifmmode {\rm L}\beta \else L$\beta$\fi}
\def  \Ha          {\ifmmode {\rm H}\alpha \else H$\alpha$\fi}
\def  \ha          {\ifmmode {\rm H}\alpha \else H$\alpha$\fi}
\def  \Hb          {\ifmmode {\rm H}\beta \else H$\beta$\fi}
\def  \hb          {\ifmmode {\rm H}\beta \else H$\beta$\fi}
\def  \Pa          {\ifmmode {\rm P}\alpha \else P$\alpha$\fi}
\def  \CIIIb       {\ifmmode {\rm C}\,{\sc iii]}\,\lambda1909
                     \else C\,{\sc iii]}\,$\lambda1909$\fi}
\def  \CIV         {\ifmmode {\rm C}\,{\sc iv}\,\lambda1549
                     \else C\,{\sc iv}\,$\lambda1549$\fi}
\def  \SIV         {\ifmmode {\rm Si}\,{\sc iv}\,\lambda1397
                     \else Si\,{\sc iv}\,$\lambda1397$\fi}
\def  \MgII         {\ifmmode {\rm Mg}\,{\sc ii}\,\lambda2798
                     \else Mg\,{\sc ii}\,$\lambda2798$\fi}
\def  \MGII         {\ifmmode {\rm Mg}\,{\sc ii}\,\lambda2798
                     \else Mg\,{\sc ii}\,$\lambda2798$\fi}
\def  \OVI         {\ifmmode {\rm O}\,{\sc vi}\,\lambda1035
                     \else O\,{\sc vi}\,$\lambda1035$\fi}

\def  \HeI        {\ifmmode {\rm He}\,{\sc i}\,\lambda5876
                     \else He\,{\sc i}\,$\lambda5876$\fi}
\def  \HeII        {\ifmmode {\rm He}\,{\sc ii}\,\lambda4686
                     \else He\,{\sc ii}\,$\lambda4686$\fi}
\def  \HeIIa        {\ifmmode {\rm He}\,{\sc ii}\,\lambda1640
                     \else He\,{\sc ii}\,$\lambda1640$\fi}
\def  \HeIIb        {\ifmmode {\rm He}\,{\sc ii}\,\lambda1085
                     \else He\,{\sc ii}\,$\lambda1085$\fi}
\def  \CII        {\ifmmode {\rm C}\,{\sc ii}\,\lambda1335
                     \else C\,{\sc ii}\,$\lambda1335$\fi}
\def  \CIII        {\ifmmode {\rm C}\,{\sc iii}\,\lambda977
                     \else C\,{\sc iii}\,$\lambda977$\fi}
\def  \CIIIb       {\ifmmode {\rm C}\,{\sc iii]}\,\lambda1909
                     \else C\,{\sc iii]}\,$\lambda1909$\fi}
\def  \CIV         {\ifmmode {\rm C}\,{\sc iv}\,\lambda1549
                     \else C\,{\sc iv}\,$\lambda1549$\fi}
\def  \bOIIIb       {\ifmmode {\rm [O}\,{\sc iii]}\,\lambda5007
                     \else [O\,{\sc iii]}\,$\lambda5007$\fi}
\def  \OIIIb       {\ifmmode {\rm O}\,{\sc iii]}\,\lambda1663
                     \else O\,{\sc iii]}\,$\lambda1663$\fi}
\def  \OVIII        {\ifmmode {\rm O}\,{\sc viii}\,653~{\rm eV}
                     \else O\,{\sc viii}\,$653~{\rm eV}$\fi}
\def  \OVII        {\ifmmode {\rm O}\,{\sc vii}\,568~eV
                     \else O\,{\sc vii}\,$568~{\rm eV}$\fi}
\def  \OVI         {\ifmmode {\rm O}\,{\sc vi}\,\lambda1035
                     \else O\,{\sc vi}\,$\lambda1035$\fi}
\def  \OIVb         {\ifmmode {\rm O}\,{\sc iv]}\,\lambda1402
                     \else O\,{\sc iv]}\,$\lambda1402$\fi}
\def  \OVb         {\ifmmode {\rm O}\,{\sc v]}\,\lambda1218
                     \else O\,{\sc v]}\,$\lambda1218$\fi}
\def  \bOIIb       {\ifmmode {\rm [O}\,{\sc ii]}\,\lambda3727
                     \else [O\,{\sc ii]}\,$\lambda3727$\fi}
\def  \bOIb       {\ifmmode {\rm [O}\,{\sc i]}\,\lambda6300
                     \else [O\,{\sc i]}\,$\lambda6300$\fi}
\def  \OI       {\ifmmode {\rm [O}\,{\sc i]}\,\lambda1304
                     \else [O\,{\sc i]}\,$\lambda1304$\fi}
\def  \NII         {\ifmmode {\rm N}\,{\sc ii}\,\lambda1084
                     \else N\,{\sc ii}\,$\lambda1084$\fi}
\def  \bNIIb         {\ifmmode {\rm [N}\,{\sc ii]}\,\lambda6584
                     \else [N\,{\sc ii]}\,$\lambda6584$\fi}
\def  \NIII         {\ifmmode {\rm N}\,{\sc iii}\,\lambda990
                     \else N\,{\sc iii}\,$\lambda990$\fi}
\def  \NIIIb         {\ifmmode {\rm N}\,{\sc iii]}\,\lambda1750
                     \else N\,{\sc iii]}\,$\lambda1750$\fi}
\def  \NIVb         {\ifmmode {\rm N}\,{\sc iv]}\,\lambda1486
                     \else N\,{\sc iv]}\,$\lambda1486$\fi}
\def  \NV          {\ifmmode {\rm N}\,{\sc v}\,\lambda1240
                     \else N\,{\sc v}\,$\lambda1240$\fi}
\def  \MgII        {\ifmmode {\rm Mg}\,{\sc ii}\,\lambda2798
                       \else Mg\,{\sc ii}\,$\lambda2798$\fi}
\def  \CVI        {\ifmmode {\rm C}\,{\sc vi}\,368~eV
                       \else C\,{\sc vi}\,368~eV\fi}
\def  \SiIV         {\ifmmode {\rm Si}\,{\sc iv}\,\lambda1397
                     \else Si\,{\sc iv}\,$\lambda1397$\fi}
\def  \bFeXb       {\ifmmode {\rm [Fe}\,{\sc x]}\,\lambda6734
                       \else [Fe\,{\sc x]}\,$\lambda6734$\fi}
\def  \MgX        {\ifmmode {\rm Mg}\,{\sc x}\,\lambda615
                       \else Mg\,{\sc x}\,$\lambda615$\fi}
\def  \MgXI        {\ifmmode {\rm Mg}\,{\sc xi}\,1.34~keV
                       \else Mg\,{\sc xi}\,1.34~keV\fi}
\def  \MgXII      {\ifmmode {\rm Mg}\,{\sc xii}\,1.47~keV
                     \else Mg\,{\sc xii}\,1.47~keV\fi}
\def  \bNeVb      {\ifmmode {\rm [Ne}\,{\sc v]}\,\lambda3426
                     \else [Ne\,{\sc v]}\,$\lambda3426$\fi}
\def  \bNeIIb      {\ifmmode {\rm [Ne}\,{\sc ii]}\,12.8\mu~m
                     \else [Ne\,{\sc ii]}\,$12.8 \mu m$\fi}
\def  \bOIVb      {\ifmmode {\rm [O}\,{\sc iv]}\,25.9\mu~m
                     \else [O\,{\sc iv]}\,$25.9 \mu m$\fi}
\def  \bNeVIRb      {\ifmmode {\rm [Ne}\,{\sc v]}\,14.3\mu~m
                     \else [Ne\,{\sc v]}\,$14.3 \mu m$\fi}
\def  \NeVIII      {\ifmmode {\rm Ne}\,{\sc viii}\,\lambda774
                     \else Ne\,{\sc viii}\,$\lambda774$\fi}
\def  \SiVIIa      {\ifmmode {\rm Si}\,{\sc vii}\,\lambda70
                     \else Si\,{\sc vii}\,$\lambda70$\fi}
\def  \NeVIIa      {\ifmmode {\rm Ne}\,{\sc vii}\,\lambda88
                     \else Ne\,{\sc vii}\,$\lambda88$\fi}
\def  \NeVIIIa      {\ifmmode {\rm Ne}\,{\sc viii}\,\lambda88
                     \else Ne\,{\sc viii}\,$\lambda88$\fi}
\def  \NeIX      {\ifmmode {\rm Ne}\,{\sc ix}\,915~eV
                     \else Ne\,{\sc ix}\,915~eV\fi}
\def  \NeX      {\ifmmode {\rm Ne}\,{\sc x}\,1.02~keV
                     \else Ne\,{\sc x}\,1.02~keV\fi}
\def  \SiXII        {\ifmmode {\rm Si}\,{\sc xii}\,\lambda506
                       \else Si\,{\sc xii}\,$\lambda506$\fi}
\def  \SiXIII      {\ifmmode {\rm Si}\,{\sc xiii}\,1.85~keV
                     \else Si\,{\sc xiii}\,1.85~keV\fi}
\def  \SiXIV      {\ifmmode {\rm Si}\,{\sc xiv}\,2.0~keV
                     \else Si\,{\sc xiv}\,2.0~keV\fi}
\def  \SXV      {\ifmmode {\rm S}\,{\sc xv}\,2.45~keV
                     \else S\,{\sc xv}\,2.45~keV\fi}
\def  \SXVI      {\ifmmode {\rm S}\,{\sc xvi}\,2.62~keV
                     \else S\,{\sc xvi}\,2.62~keV\fi}
\def  \ArXVII      {\ifmmode {\rm Ar}\,{\sc xvii}\,3.10~keV
                     \else Ar\,{\sc xvii}\,3.10~keV\fi}
\def  \ArXVIII      {\ifmmode {\rm Ar}\,{\sc xviii}\,3.30~keV
                     \else Ar\,{\sc xviii}\,3.30~keV\fi}
\def  \FeI_XVI      {\ifmmode {\rm Fe}\,{\sc 1-16}\,6.4~keV
                     \else Fe\,{\sc 1-16}\,6.4~keV\fi}
\def  \FeXVII_XXIII     {\ifmmode {\rm Fe}\,{\sc 17-23}\,6.5~keV
                     \else Fe\,{\sc 17-23}\,6.5~keV\fi}
\def  \FeXXV      {\ifmmode {\rm Fe}\,{\sc xxv}\,6.7~keV
                     \else Fe\,{\sc xxv}\,6.7~keV\fi}
\def  \FeXXVI      {\ifmmode {\rm Fe}\,{\sc xxvi}\,6.96~keV
                     \else Fe\,{\sc xxvi}\,6.96~keV\fi}
\def  \FeLa     {\ifmmode {\rm Fe}\,{\sc L}\,0.7-0.8~keV
                     \else Fe\,{\sc L}\,0.7-0.8~keV\fi}
\def  \FeLb     {\ifmmode {\rm Fe}\,{\sc L}\,1.03-1.15~keV
                     \else Fe\,{\sc L}\,1.03-1.15~keV\fi}
\def\hi{\ifmmode \mbox{{\rm H}\,{\sc i}} \else H\,{\sc i}\fi}

\def\heii{He\,{\sc ii}}

\def\ciii{\ifmmode {\rm C}\,{\sc iii]} \else C\,{\sc iii]}\fi}
\def\civ{\ifmmode {\rm C}\,{\sc iv} \else C\,{\sc iv}\fi}
\def\cv{\ifmmode {\rm C}\,{\sc v} \else C\,{\sc v}\fi}

\def \XMM    {{\it XMM}}

\def \swift {{\it Swift}}
\def \hst   {{\it HST}}
\def \HST   {{\it HST}}

\def  \L46         {$ L_{46} $}

\def  \Ledd        {$ L/L_{\rm Edd} $}    
\def  \MBH         {$ M_{\rm BH} $}     
\def  \M*         {$ M_{*} $}     
\def  \m9          {$ m_9 $}
\def  \N10         {$ N_{10} $}

\title[Continuum lags in AGN]{Continuum reverberation mapping and a new lag-luminosity 
relationship for AGN}

\author[Hagai Netzer]{
Hagai Netzer,$^{1}$\thanks{E-mail: \url{hagainetzer@gmail.com}}
\\
$^{1}$School of Physics and Astronomy, Tel Aviv University, Tel Aviv 69978, Israel \\
}

\date{Accepted XXX. Received YYY; in original form ZZZ}

\pubyear{2021}

\hypersetup{draft}
\begin{document}
\label{firstpage}
\pagerange{\pageref{firstpage}--\pageref{lastpage}}
\maketitle






\begin{abstract}

High cadence, high quality observations of active galactic nuclei (AGN) clearly show continuum variations
with lags, relative to the shortest observed variable UV  continuum,  that increase with wavelength (``lag spectra'').
These have been attributed to the irradiation and heating of the central
accretion disk by the central X-ray emitting corona. An alternative explanation, connecting the observed
lag-spectra  to line and continuum emission from gas in the broad line region (BLR), has also been proposed. 
In this paper I show the clear spectral signature of the time-dependent diffuse gas emission in the lag-spectrum of 6 AGN. 
I also show a new lag-luminosity relationship for 9 objects which is a scaled down version
of the well known $\tau(\hb)$-\Lop\ relationship in AGN.
The shape of the lag-spectrum, and its normalization, are entirely consisted with diffuse emission from radiation pressure
supported clouds in a BLR with a covering factor of about 0.2. While some contributions to the continuum lag from
the irradiated disk cannot be excluded,
there is no need for this explanation.

\end{abstract}


\section{Introduction}

Super-massive black holes (BHs), central accretion disks, and strong broad emission lines originating
from high density gas in the broad line region (BLR), are some of the main characteristics of type-I active galactic nuclei (AGN). 
 The spectral energy distribution (SED) of the accretion disk, and the line and continua emitted by the BLR gas, all show
 variations on time scales of days-to-years, depending on the bolometric luminosity of
 the source (\LAGN). A highly variable central X-ray source, thought to be produced by  a hot corona in the vicinity
 of the BH, is another characteristic signature of such objects.
 The spectra of such sources have been studied in great detail  for years. This includes hundreds of thousands of sources at all redshifts, from zero to beyond 7 \citep[see review and numerous references in][]{
 Netzer2013}.

Progress in reverberation mapping (RM) of the BLR gas provides an opportunity to map the location and motion of the gas, and to estimate the BH mass (\MBH). 
Such measurements are available for a large number of sources and for several broad emission lines 
\citep[e.g.][]{Kaspi2000,Bentz2010,Bentz2013,Hu2015,Du2015,Grier2017,Pei2017,Lira2018,Kriss2019,Bentz2021}.
They have been used to derive observables such as ``the mean emissivity radius'' or 
``the mean responsivity radius'', and to show that these radii are proportional to  
to $L_{5100}^{1/2}$, where \Lop\ is the monochromatic continuum luminosity at 5100\AA\  in \ergs. 
For sources in the range $10^{44}<$\LAGN$<10^{46}$\,\ergs, both the mean emissivity and the mean responsivity radii
 of the \hb\ line, $r($\hb), are about 34\Lopna\ light days (ld), 
 where \Lopn\ refers to the value of \Lop/$10^{44}$\,\ergs\
 \citep[e.g.][]{Kaspi2000,Bentz2013}.
 
K-band interferometry \citep{gravity2020}, and combined V-band and K-band photometry, \citep[e.g.][]{Koshida2014}, provide information about the outer boundary of the BLR. 
This boundary is roughly at the  sublimation radius of graphite grains which is  $\sim 3 \times r(\hb)$ 
  \cite[e.g.][and references therein]{Netzer2015}.
 Detailed theoretical studies, like \cite{Baskin2018},  show that a single, well defined outer boundary  is a simplified description of a more complex configuration connecting the inner accretion disk to an outer,
 torus-like structure.
 
 More recent RM studies 
 follow the optical-UV continuum variations relative to the central X-ray source 
 \citep[so-called ``continuum RM'', see][and references therein]{Fausnaugh2016,Edelson2019,Cackett2020}. Such variations are predicted to take place in sources with large X-ray luminosity (\Lx) due to the
  illumination and heating of the central accretion disk by  radiation from the hot
 corona. Simple irradiated disk models
 predict time delays which are proportional to $\lambda^{4/3}$, where the wavelength $\lambda$ is used as a proxy
 for the surface temperature of the disk. More advanced models, taking into account the transfer of the X-ray signal
 across the disk and relativistic ray bending,  result in much longer predicted lags  \citep{Kammoun2021a,Kammoun2021b}. Whether or not such models can explain the observations is still an open question.
 In some sources, the optical-UV variation are, indeed, following the X-ray variations. However, in several others this is not the case and the lag is 
 either ill-defined, not following the predicted $\lambda^{4/3}$ dependence, or the UV leads the X-ray in contrast to the prediction of the model 
   \citep[e.g.][]{Edelson2019,Cackett2020}. 
   
   Alternative explanations to the observed continuum RM involve time-variable emission from the diffuse BLR 
   gas. The variations are the results of the time-variable ionizing luminosity emitted 
   by the disk which induce continuum variations that are not related to
  X-ray irradiated. The effects of the time dependent bound-free and free-free continuum emission have first been studied by \cite{Korista2001} and in greater detail by  
  \cite{Lawther2018}, \cite{Chelouche2019} and \cite{Korista2019}. Detailed comparison  of the model prediction to the observations of several sources 
  are still missing and there are several open issues related to the size of the lags and the comparison with broad emission line variations observed in the same BLR.

In this paper I present detailed evidence that most, and perhaps all of the continuum RM observed in AGN are due to
the time-dependent emission of the diffuse BLR gas. This is done by comparing lag-spectra of 9 AGN with the results of radiation pressure confined (RPC) cloud models. I also present a new lag-luminosity relationship for AGN and suggest ways to use it as a test of the model.  
In \S2 I describe the sample. In \S3 I present predicted irradiated disk lag-spectra and contrast them with the observations.
In \S4 I discuss diffuse emission and dust
lag spectra and present a new lag-luminosity relationship for AGN. \S5 gives a discussion of the new results and \S6 presents the conclusions of the paper. 
Throughout this paper I assume a standard cosmological model with $\Omega_{\Lambda}=0.7$, $\Omega_{m}=0.3$, and
$H_{0}=70\,\kms\,\, {\rm Mpc}^{-1}$.

\section{Sample and data}

The sample used here contains 9 AGN with high quality broad band observations from the hard X-ray to the near infrared (NIR). 
The main facilities used are \XMM, \swift, \HST, and a number of ground based telescopes, including several 1-2m telescopes of the LCO observatory. 
The observations are described and discussed in a large number of papers including \cite{Fausnaugh2016}, \cite{Fausnaugh2018}, \cite{Edelson2019}, \cite{Cackett2020} and 
\cite{Kara2021}.

The 9 objects in the sample  are separated into two groups.
Group A includes 6 AGN with high quality, high cadence observations. These are: NGC\,5548, NGC\,4593, NGC\,2617, F9, Mrk\,142 and Mrk\,817. Group B includes 3 objects with lower quality data which are still good enough to obtain reliable lags for both  \hb\  and the 5000-5500\AA\ continuum. 
The object are NGC\,4151, Mrk\,509, and NGC\,7469. 

Table~1 provides basic information about all the sources, including \MBH, \Lop, \LAGN,
the \hb\ lag ($\tau(\hb)$), the 5000-5500\AA\ lag (hereafter $\tau_{5100,cont}$), and the references used to obtain
this information.  The bolometric luminosity, \LAGN, was obtained by applying the disk bolometric correction factor 
proposed by \cite{Netzer2019} to the observed \Lop. This may differ, slightly,
from estimates that are based on other methods. As for the X-ray luminosity, I assume that the sole energy source is accretion through an optically thin, geometrically thick accretion disk 
\citep{Shakura1973}. Therefore, \Ldisk=\LAGN\ and \LX\ cannot exceed a certain
fraction of \Ldisk\ which is assumed to be 0.5. I also assume  $L_{\rm Edd}=1.5 \times 10^{38}$(\MBH/\Msun).

Several additional comments are in order:
\begin{enumerate}
 \item 
All objects are highly variable on time scales of months-to-years with variability amplitude as large as a factor two at some wavelengths, mostly in the UV band. The typical variation at 5100\AA\ is of order 30\% and the typical uncertainties on the numbers shown in Table 1 are of this order. 
		Moreover, most of the objects in this small sample, which includes all AGN with high quality continuum
		RM observations at the time of writing, show $\tau(\hb)$ which falls below the \cite{Bentz2013}
		relationship by up to a factor two. Thus, the mean \hb\ lag is not a fair representation of the entire population. 

In several cases, e.g. NGC\,4151, the listed $\tau(\hb)$ and $\tau_{5100,cont}$
 were observed in different campaigns and there is clear evidence for a large change in \Lop\ between
 the two epochs. I have adjusted \Lop\ in such cases to the most recent campaign. The uncertainty on the
 $\tau(\hb)/\tau_{5100,cont}$ ratio in such cases is larger than the one found by a simple combination of the two uncertainties.  
\item
 I do not discuss the problematic issue of the UV-optical lag
relative to the X-ray flux (\S1). I simply assume that the distance 
 of the X-ray illuminating source from the central BH is very small, a fraction of a light day.
\item
Given point (ii), there is no reference point to the measured lags, only inter-band time-delay (IBTD) measurements. The shortest wavelength band
observed by \HST, at around 1150\AA,   
provides the best reference point since it is close in wavelength to both the  ionizing and illuminating continua.
When available, all lags are measured relative to this band. Otherwise, I use the \swift\ UVW2 band centered at around 1928\AA\ and note 
 a clear positive lag between 1150\AA\ and this band, by typically 0.5-1 days. The difference in lag seems to follow both the predictions of  the standard illuminated-disk model and the new model  presented below  
		hence I used these calculations to derive  a ``zero lag'' point at 1200\AA.
		This approach is complicated due to the large uncertainties on the Rayleigh scattering feature \citep{Korista1998} which is peaked close to the \La\ line center (see further details in \S~\ref{section:comparison_detailed_models}). Finally, I did not change the observed $\tau(\hb)$ which is usually measured 
		relative to the 5100\AA\ continuum. Because of the continuum RM, the lag of this line relative to the driving UV continuum is longer than listed in table~1. 

\end{enumerate}

\onecolumn

   

\begin{table*}
\centering
\caption{Sample properties}
\begin{tabular}{lcccccccc}
Source & D$_{\rm L}$ & \MBH\ & \Lop\   & \LAGN$^a$  & $\tau(5000-5500)^c$ & $\tau(\hb)^c$ &  $f_{irr}^b$ & References \\ 
       &      Mpc   & \Msun\ &\ergs\  & \ergs\  &    days                      &  days       &            & \\
\hline
Group A   &&&&&&&& \\
\hline
NGC\,5548 &74.5  & 5e7   &2.3e43 & 2.8e44 & 2.16&10  & 0.8& 1,2 \\
NGC\,4593 & 38.8 & 7.6e6 & 7.4e42 & 7.4e43 & 0.57& 3.7& 0.8 &1,5  \\
NGC\,2617 & 61.5 &3.2e7  & 5.5e42 & 1.0e44 & 0.75& 4.3& 0.8 & 1,3,10 \\
Mrk\,142  & 199  & 1.7e6 & 3.5e43 & 6.8e44 & 0.8 &6.0 & 0.08 & 1,6 \\
F9        & 203  & 2.6e8 &7.9e43 & 1.3e45 & 4.0 & 18 & 0.28 & 1,9 \\
Mrk\,817  &134.2 & 3.9e7 &6.9e43 & 1.1e45 & 4.2 & 22 & 0.15 & 1,4\\
 &&&&&&&& \\
Group B   &&&&&&&& \\
\hline
NGC\,4151  &16.6 & 3.6e7 &3.2e42 & 5.4e43 & 1.4& 8  &0.8& 1,7 \\
NGC\,7469  &70.8 & 9.0e6 &3.6e43 & 7.1e44 & 1.7& 11 &0.19 &1,8  \\
Mrk\,509   &151  & 1.1e8 &1.4e44 & 8.4e44 & 3.3& 77 & 0.8& 1,7 \\
\hline
\end{tabular} 
\begin{tablenotes}
 \item
 $^a$ Calculated from \Lop.
 $^b$ Calculated in eqn.~\ref{firr} assuming $\eta=0.1$ and requiring $f_x \le 0.5$.
 $^c$ Uncertainties can be found in the original papers.
 \item
 $^1$ The AGN BH Mass Database ~http://www.astro.gsu.edu/AGNmass/
 $^2$ \cite{Fausnaugh2016}
 $^3$ \cite{Fausnaugh2018}
 $^4$ \cite{Kara2021}
 $^5$ \cite{Cackett2018}
 $^6$ \cite{Cackett2020}
 $^7$ \cite{Edelson2019}
 $^8$ \cite{Pahari2020}
 $^9$ \cite{Hernandez2020}
 $^{10}$ \cite{Fausnaugh2017}

\end{tablenotes}

\label{tab:table1}
\end{table*}

\twocolumn

\section{Accretion disks and observed lag-spectra}

\subsection{Theoretical disk lag spectra}

The accretion disk model used to calculate the SED is the one presented in  \cite{Slone2012}. It  includes relativistic corrections and Comptonization in the disk atmosphere. This is rather standard and other codes used in the literature result in similar SEDs.
The total accretion energy is \Ldisk$=\eta \dot{M} c^2$ where $\eta$ is the mass-to-energy conversion efficiency which changes between  0.0398 ($a=-1$) to 0.328 ($a=0.998$), where $a$ is the spin parameter. Part of the accreted energy is converted to  X-ray powerlaw radiation  in a hot corona. The only X-ray photon index considered here is $\Gamma=2$. This is close
to the mean of the present sample that range between 
1.4 and 2.4.

The observed lags were compared with two predicted irradiated disk lags.
The first is the standard equation \citep[e.g.][]{Fausnaugh2016} presented here as:
\begin{equation}
	\tau_{\lambda, irr}=1.62 \times 10^{-5} \lambda_A^{4/3} \left[ \frac{X}{4.97} \right] ^{4/3} [ M_7 \dot{M_{\odot} } ]^{1/3} [3+f_{irr} ]^{1/3} \,\, {\rm days },
	\label{disk_standard}
\end{equation}
where $M_7$ is the BH mass in units of $10^7$ \Msun, $\dot{M_{\odot} }$ is the accretion rate in units of \msun/yr, and
$\lambda_A$ is the wavelength in Angstroms. The scaling parameter $X$ is 4.97 for a simple Wien's law 
and $\approx 2.49$ when averaging contributions from the entire surface of the disk. In all cases shown
here I use $X=2.49$ and note that this, by itself, introduces a factor of about 2.5 uncertainty on $\tau_{\lambda}$.

The term 
$f_{irr} $ represents the irradiated X-ray flux absorbed by the disk. It is given by:
\begin{equation}
f_{irr} = 2 (1-al) f_x \eta h  \,\, ,
\label{firr}
\end{equation}
where $al$ is the albedo of the disk atmosphere,  
$f_x=$\Lx/\Ldisk\ is the fraction of the total produced power emitted  as X-ray radiation, and $h$ is the height of the X-ray source above the 
plane of the disk in units of $r_g=GM_{BH}/c^2$. 
$f_{irr}$ is basically wavelength independent except for cases where $h$ is very large. In what follows I assume
$al=0.2$, $\eta=0.1$, and $h=10$.  

The total corona-produced emission is a question of much debate. It depends on low (unknown) and high (known in many cases) cut-off energies. Here I use $E_{max}=100$ keV, as observed in many cases, and $E_{min}$ which is model-dependent and is probably in the range of 50-500 eV. \cite{Kammoun2021a} 
discussed this issue at great length and I used their eqn.~5 to estimate the total X-ray energy, \Lx. In some of the sources the resulting  $E_{min}$ is very small, 1 Rydberg or even lower. This can result in $L_X \sim $\Ldisk\ which is inconsistent with the basic energy production assumption. I therefore require $f_x \le  0.5$\Ldisk. 
 Values of
$f_{irr}$ are listed in table~1.

I have also compared the lags with the predictions of the \cite{Kammoun2021a} model. Such fits are   
 not discussed here  since they were shown in the original papers.

\subsection{Observed lag spectra}

\begin{figure*} \centering
        \includegraphics[width=0.45\linewidth]{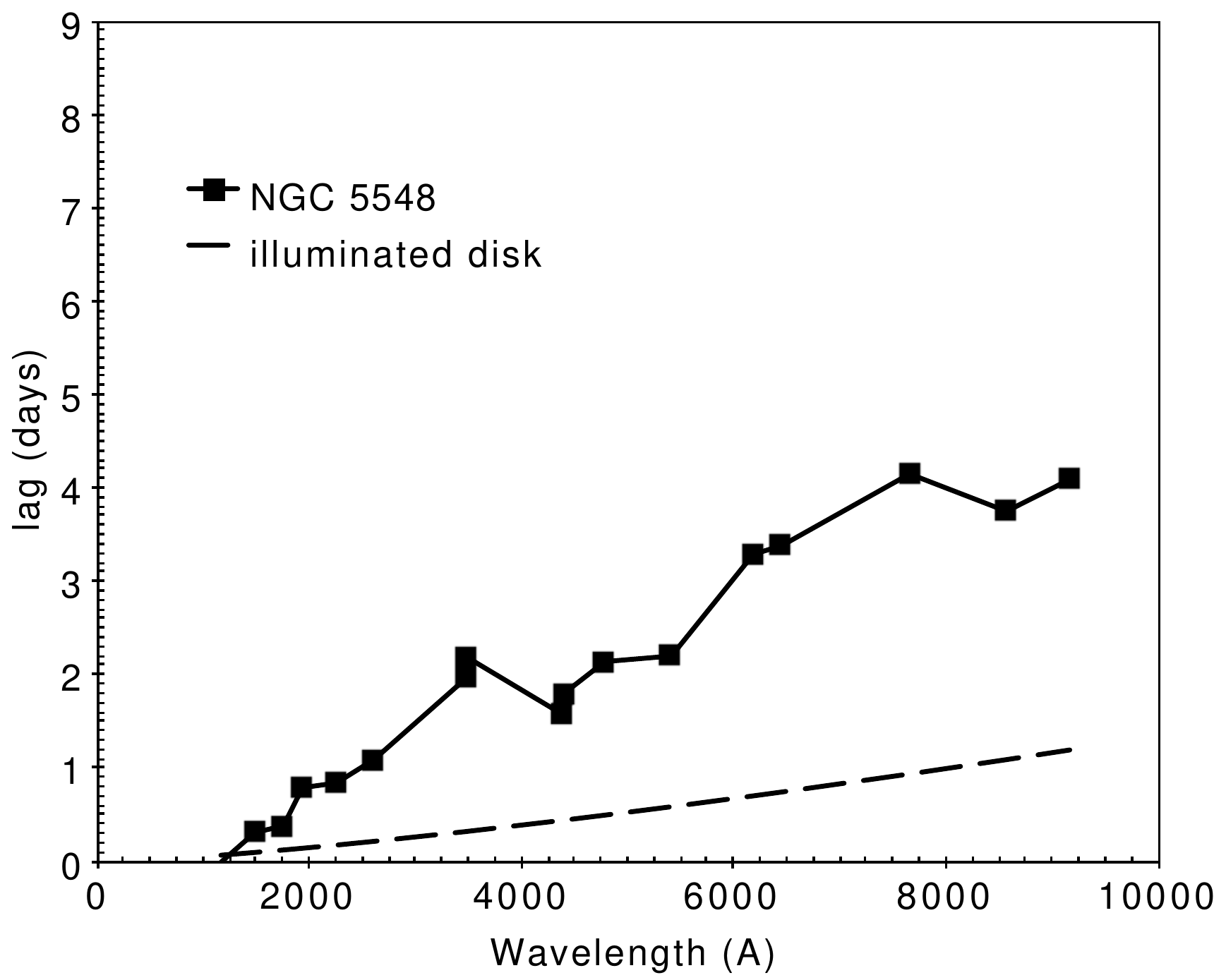}
        \includegraphics[width=0.45\linewidth]{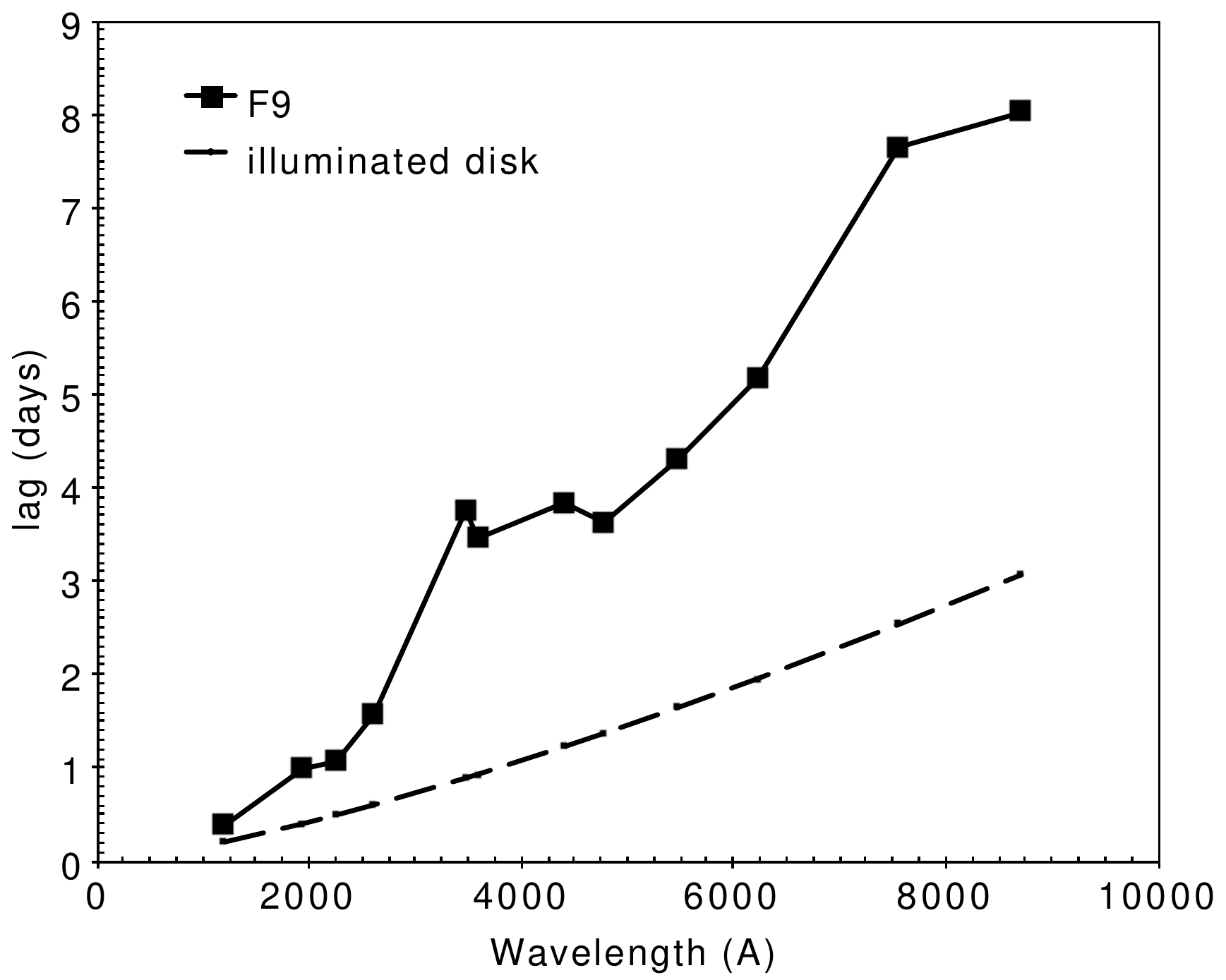}\\
        \includegraphics[width=0.45\linewidth]{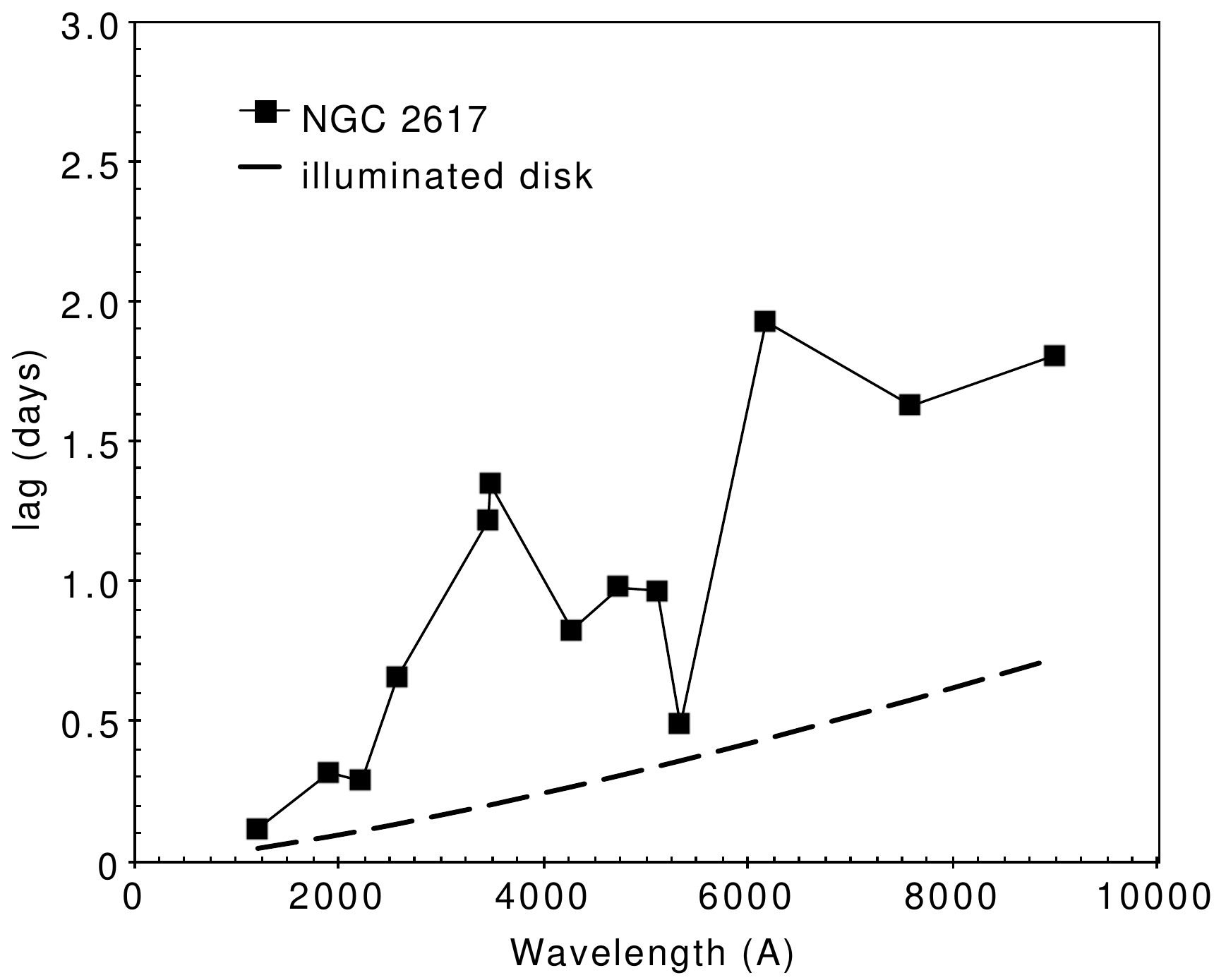}
        \includegraphics[width=0.45\linewidth]{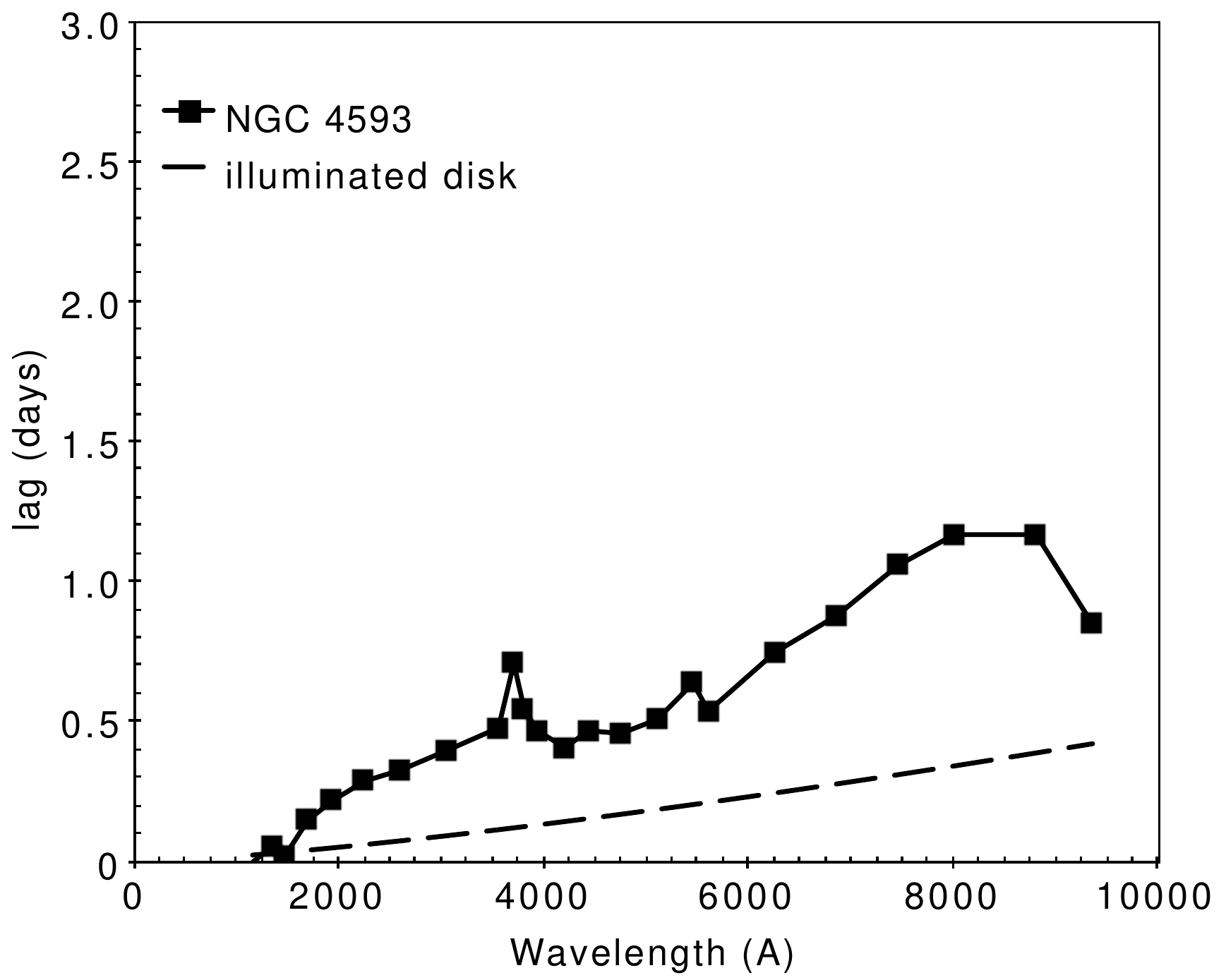}

\caption{Lag spectra for four sources with high quality observations.    
Note two clear bumps in all lag-spectra resembling real spectral features in many AGN.
The 2000-4000\AA\ feature is similar in shape and in wavelength to the ``small blue bump'' observed
in most type-I AGN and the peak of the 5000-10000\AA\ feature is close to the wavelength of the Paschen jump at 8200\AA.
} 
\label{central_figure}
\end{figure*}

Fig.~\ref{central_figure} shows  observed lag-spectra  of 4 group-A AGN. I compare
them with eqn.~\ref{disk_standard} using
the parameters listed in table~1. As explained, the observed
lags were shifted upward by a constant obtained from the extrapolation of eqn.~\ref{disk_standard} to 1200~\AA\ to achieve a zero lag at this wavelength.
The uncertainly introduced by this adjustment is small compared with the measured continuum lags and makes 
little difference to the conclusion of the paper. 
I do not plot the lag uncertainties since they obscure the spectral features that are at the center of this paper.

All lag spectra in Fig.~\ref{central_figure} clearly show two distinct bumps with maxima at around 3500\AA\ and $\sim$ 8000\AA, and a local minimum at around 5000\AA.
The shorter wavelength bump resembles the commonly observed ``small blue bump'' in the spectrum of most type-I AGN.
The existence of a time-lag excess in the $u$ or $U$ bands was noted in earlier publications \citep[e.g.][]{Edelson2019,Cackett2018}. It was interpreted as due mostly to the Balmer jump
	at 3646\AA.  Closer examination shows that the feature is much broader and common to {\it all} type-I AGN with high quality lag observations. 
	Contributing to the feature, on
	top of the Balmer continuum emission, are strong broad FeII lines on the short wavelength side and a large number of high order Balmer lines that extend the 
	continuum-looking emission to longer wavelengths all the way to about 4000\AA, depending on the source \citep[see e.g.][]{Wills1985,Mejia2016}.

	The second bump covers the 5000-9000\AA\ wavelength range. 
	The rise at shorter wavelengths and drop at longer wavelength are clearly visible and the peak at around 7500-8500\AA\ is very noticeable
	suggesting a combination of Paschen jump in emission at around 8200\AA, Paschen continuum drop towards $\sim 5000$\AA,
	and the conglomeration of many broad Paschen lines longward of the jump. The broad band photometry includes, in some of the sources,
	 at least part of the strong and broad \Ha\ line which also contributes to this feature. This is not the case in NGC\,4593 where the data used purposely avoid this and other strong lines. 
	 
	The lags due to simple disk irradiation are small in all cases and 
	contribute little, if anything, to the lag-spectrum. The \cite{Kammoun2021b} lags, which are not shown here,  are similar in shape but
	longer by about a factor 2. None of the irradiated models explains 
	the two humps observed in the lag-spectra.

\section{Diffuse emission and dust Lag spectra}

The combined luminosity of the accretion disk and the central X-ray source are marked here by $L_{incident}$.
I also use two different symbols to describe the luminosity produced by the BLR gas. $L_{diff}$ stands for the combined luminosity of the broad emission lines,  
bound-free and free-free continua, and scattering by ionized and neutral gas. For the diffuse continuum (DC), I use $L_{DC}$. Thus 
$L_{total}=L_{diff}+L_{incident}$. Emission by hot dust close to the center is counted separately as $L_{dust}$. All these luminosities are wavelength dependent.

\subsection{Diffuse BLR emission}\label{section:diffuse_BLR_emission}

In \cite{Netzer2020} I presented extensive {\it Cloudy} and {\it ION} photoionization calculations of BLR clouds whose structure was determined by the radiation pressure force of the central source (radiation pressure confined clouds, RPC, see \cite{Baskin2018}).
The calculations were applied to a source referred to as a ``typical RM sample'' AGN, and to NGC~5548.  For the RM sample source, there were two SEDs which assumed \MBH=$10^8$\,\Msun\ and two different accretion rates. These were combined with typical X-ray powerlaw continua. Distance from the BH, ionizing flux and diffuse emission, were all normalized to \LAGN\ which, in itself, was normalized to \Lop. The computations can directly be applied to the present sample by scaling to the observed \Lopn. Unless otherwise stated, the SED used in the present paper is the one presented in \cite{Netzer2020} as AD1.

The \cite{Netzer2020} model resulted in calculated lags of several emission lines
assumed to correspond to the emissivity weighted radii of the lines. This is a good approximation when the time scale for the ionizing continuum variations is longer than the crossing time of the BLR. The most important parameter of the model, which determines the line luminosity and 
 lag, is the distant dependent covering factor, $c_f(r)$. 
For comparison with the observations of the Balmer lines I used the population mean $\tau(\hb)$  \citep{Bentz2013} expressed in a somewhat different form as,
\begin{equation}
 \tau(\hb) = 34  L_{5100,44}^{1/2} \,\, {\rm days} \,\, ,
 \label{rm_hb}
\end{equation}

The model was also used to calculate the emissivity weighted radii, and therefore the lags of the integrated Balmer and Paschen continua that dominate the 2000-9000\AA\ range. For the chosen SED, this was found to be,
\begin{equation}
 \tau_{DC} \approx 17  L_{5100,44}^{1/2} \,\, {\rm days} \,\,.
\label{rm_balmer}
 \end{equation}
A potential deviation at short wavelength is a strong Rayleigh scattering bump centered around \La. The intensity of this radiation was investigated by \cite{Korista1998}. It is proportional to the column density of neutral hydrogen which, in the RPC model, is not well defined. This part of the spectrum is briefly discussed in 
\S~\ref{section:comparison_detailed_models}.

Variable ionizing radiation results in time and wavelength dependent diffuse emission, $L_{diff}$, and a corresponding diffuse emission lag, $\tau_{\lambda,diff}$, relative to the ionizing continuum light curve.  Here I investigate the situation where the diffused emission and the incident continuum light curves cannot be separated and appear as a single combined light curve. This light curve is then cross-correlated with the incident continuum light curve. The computed lag depends, therefore, on the relative scaling of the two components. 

In this paper I follow earlier studies 
\citep[e.g.][]{Korista2001,Lawther2018} and assume that the lag of the combined diffuse-incident light curve relative to the incident continuum light curve is proportional to $\tau_{\lambda,diff}$ and the proportionality factor is       
$L_{diff}/L_{total}$. This is easy to justify if there is a 1:1 correspondence between $L_{diff}$ and $L_{incident}$ across the entire BLR. However, for every line or diffuse continuum, there are parts of the BLR where this is a good approximation and others where it is not.

A related complication is the way used to measure lags from the peak, or centroid,  of the cross correlation function (CCF). The location of the peak depends on the shapes of the driving and  the combined light curves
which can introduce deviations from a simple scaling with  $L_{diff}/L_{total}$. 

These issues were 
discussed in \cite{Korista2019} who analyzed three different campaigns of NGC\,5548 using the LOC framework. They showed (see \S2.6\ and figure 11 in their
paper) that deviations
by factors of up to 2-3, occur in their model in cases where $L_{diff}/L_{total}$ is far from 0 or 1. 

  The RPC model of \cite{Netzer2020}  is different, considerably, from
the constant gas density and the constant ionization parameter models of \cite{Lawther2018}, and from the LOC model of \cite{Korista2019}. Thus, the lags predicted below can differ, substantially, from the results presented in those papers. 
In an appendix to this paper I show that the   
wavelength dependencies of the DC emission in the RPC model is rather similar across the BLR and deviations from a 1:1 relation are not necessarily large. This, however, is not a replacement for a full investigation of the issue of linearity in a general and global BLR model. Such an investigation is beyond the scope of the present paper.

For the remaining of the paper I use the simple approximation described above. I also add the disk illumination term by applying a similar scaling factor. This results in a total lag of:  
\begin{equation}
 \tau_{\lambda,tot} =
 \tau_{\lambda,irr}  \left[ \frac{L_{incident}}{L_{total}} \right]
  + 
  \tau_{\lambda,diff} \left[ \frac {L_{diff}}{L_{total}} \right] \,\, {\rm days} 
 \label{tau_total}
\end{equation}
Unless otherwise specified, I replace $\tau_{\lambda,diff}$ by $\tau_{DC}$ from
eqn.~\ref{rm_balmer} which means that wavelength dependencies are entirely due to $L_{diff}/L_{total}$. 
The above expression is similar, but not identical to eqn.~22 in \cite{Lawther2018}.

\begin{figure} \centering
        \includegraphics[width=0.95\linewidth]{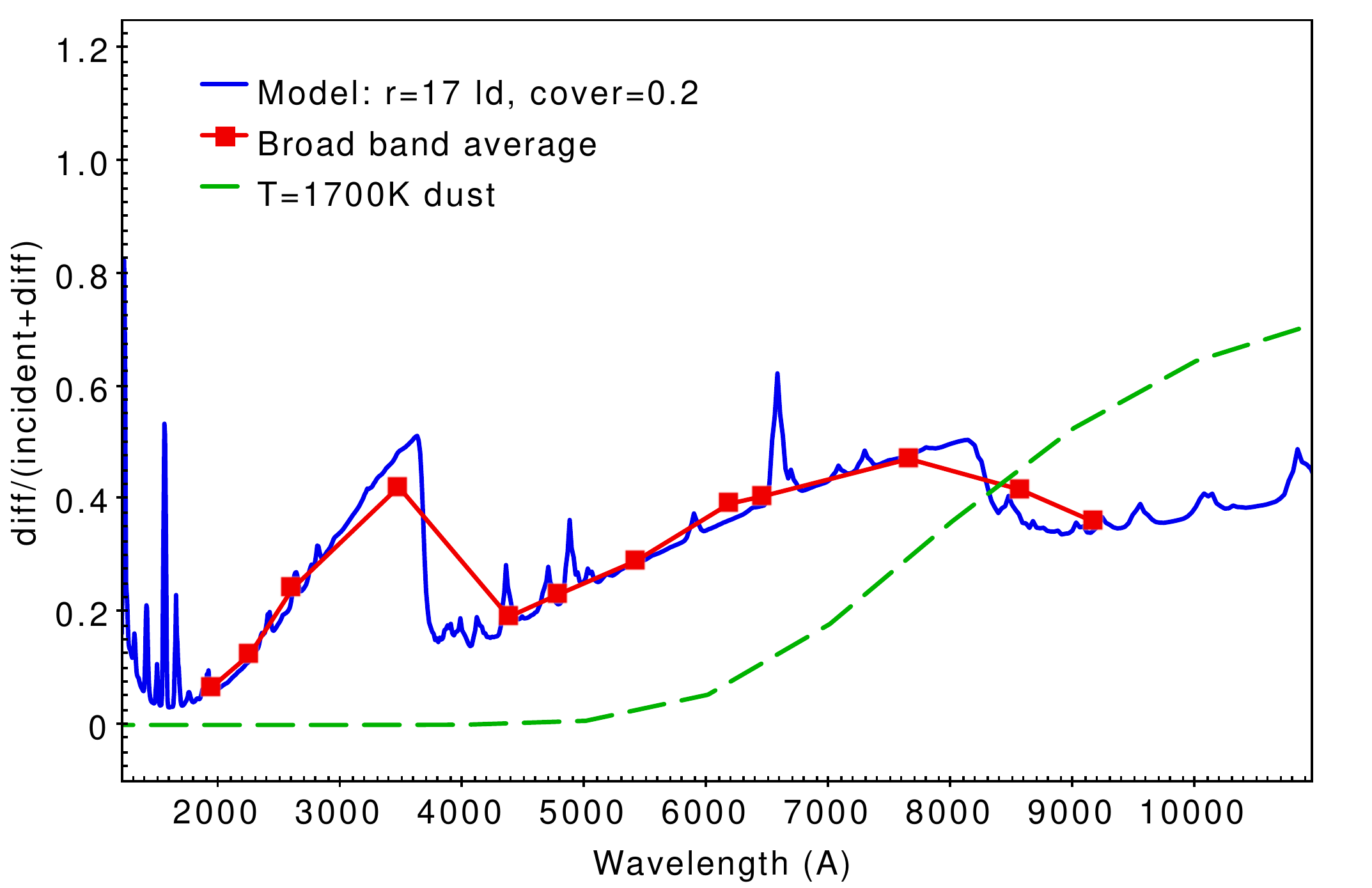}
\caption{Calculated lag spectrum (blue), and averages in several broad bands (red), for a standard RPC cloud situated 
$17 \times L_{5100,44}^{1/2}$~ld from the central BH (see text for details). The green dashed line is the contribution of the hot torus dust normalized to the mean observed
	\Lop/$L_{2 \mu m}$.
}
\label{diff_over_total}
\end{figure}

Fig.~\ref{diff_over_total} shows calculations of $L_{diff}/L_{total}$ obtained from the \cite{Netzer2020} model
assuming a single 
RPC cloud with $c_f=0.2$ situated
at the mean emissivity radius of the Balmer and Paschen continua (eqn.~\ref{rm_balmer}). 
This distance was chosen since DC emission is the largest
contributers to $L_{diff}$ at most distances. In the appendix to this paper I show that the differences between this approximation and the full BLR model are small. The broad band photometry of most of the objects in the present sample include also contributions from emission lines, mostly FeII lines on the short wavelength side of the 2000-4000\AA\ bump and several Balmer lines. The emissivity weighted 
radii of the FeII lines are 2-3 times larger than the distance of the Balmer and Paschen continua.  The relative contributions of these features to the measured lags are discussed in \S5.


In what follows I use $\tau_{5100,cont}$ to refer to measurements over the range of 5000-5500\AA. $L_{diff,5100}$  refers to the BLR emission over the same wavelength range. This band includes 
 broad FeII lines that are very weak in the present calculations
mostly because of the chosen distance.
According to Fig.~\ref{diff_over_total},  
$L_{diff,5100}/L_{total} \approx 0.23$ for \Lopn=1 and $c_f=0.2$. Since $L_{diff} \propto c_f$, and since at this wavelength $L_{diff}\approx L_{DC}$,
we can combine the calculated spectrum  with eqns.~\ref{rm_balmer} to obtain: 
\begin{equation}
 \tau_{5100,cont} \approx 17 L_{5100,44}^{1/2} \frac{1.5  c_f}{1+1.5 c_f} \,\, {\rm days} \,\, .
 \label{rm_5100}
\end{equation}
For \Lop=$10^{44}$~\ergs, $\tau_{5100,cont} \approx 3.9$ days.
Fig.~\ref{diff_over_total} can be used, in exactly the same way, to calculate the second term in eqn.~\ref{tau_total} for all other wavelengths.

The constants in all these equations are based on observations that do not take into account the diffuse flux at 5100\AA\ and the
fact that the \hb\ lag is measured relative to the 5100\AA\ continuum which, in itself, lags the short wavelength continuum. Taking this
into account would change the constants in eqns.~\ref{rm_hb} and \ref{rm_balmer} by about 12\%. The calculations presented below do not include this correction.

\subsection{The continuum lag vs. \Lop\ relationship}

I have tested the model predictions in two related ways: a comparison with observed lag-spectra of individual sources, and a test of the predicted correlation
between $\tau_{5100,cont}$ and \Lop\ for the entire sample.

\begin{figure*} \centering
        \includegraphics[width=0.45\linewidth]{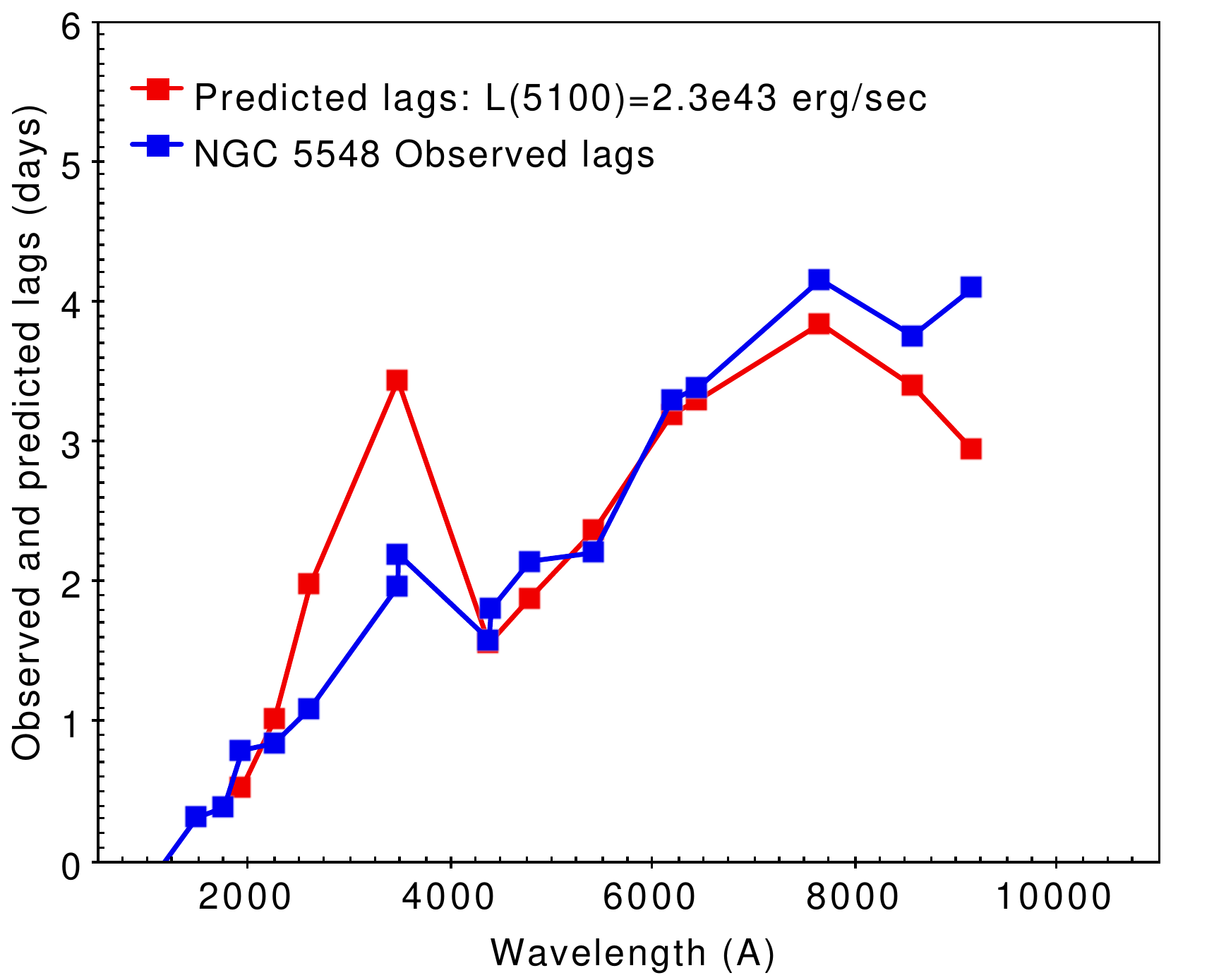}
        \includegraphics[width=0.45\linewidth]{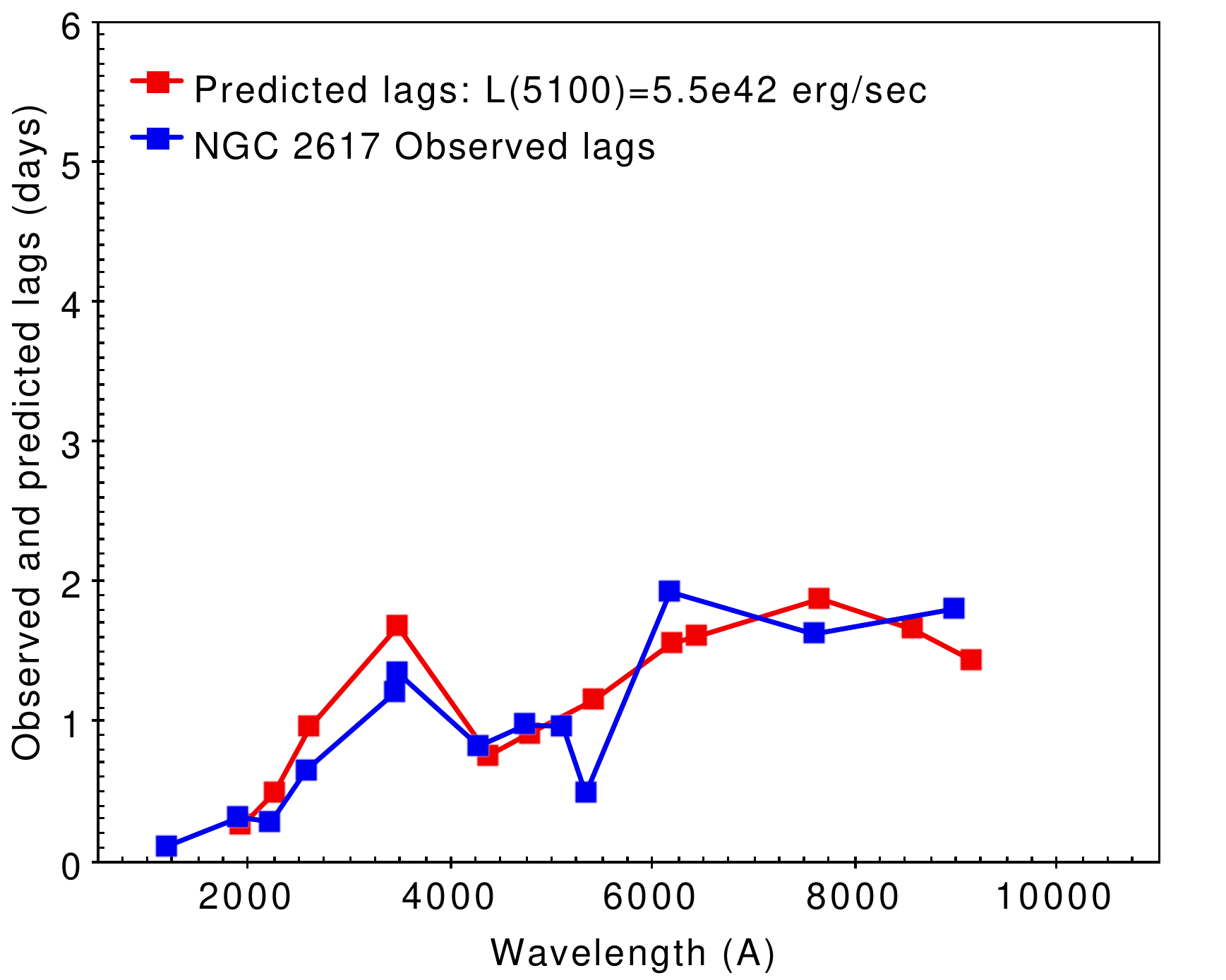}
\caption{Observed and calculated broad-band lag-spectra for NGC~5548 and NGC~2617. The model lag-spectrum in both cases was scaled by the observed \Lop\ according
to eqn.~\ref{rm_5100} assuming $c_f=0.2$. 
}
\label{n5548_n2617}

\end{figure*}

 I used the transmission curves shown in \cite{Fausnaugh2016}, together with the \swift\ bands UVW2, UVM2 and UVW1,
to calculate broad-band time lags using eqn.~\ref{tau_total} and the spectrum shown in Fig.~\ref{diff_over_total}, and assuming $\tau_{\lambda,irr}=0$.
Specific examples are shown in Fig.~\ref{n5548_n2617} where I compare model predictions with the
observations of NGC~5548 and NGC~2617. The only scaling of the model is by the observed \Lop\ of the two sources.
The agreement between model and observations is very good suggesting that the entire lag spectra of these sources, in shape and in amplitude,
are due to the variable diffuse emission.
It also suggests that the mean covering factor of the BLR in these sources is close to 0.2. 
Additional tests, not shown here, suggest similar agreement for Mrk~817 and F9. In the case of NGC~4593, the shape of the lag-spectrum is similar but $c_f=0.1$ gives  a better agreement to the observations. This is also the  case with Mrk~142 although this source is somewhat
different as  discussed below.

\begin{figure} \centering
        \includegraphics[width=0.95\linewidth]{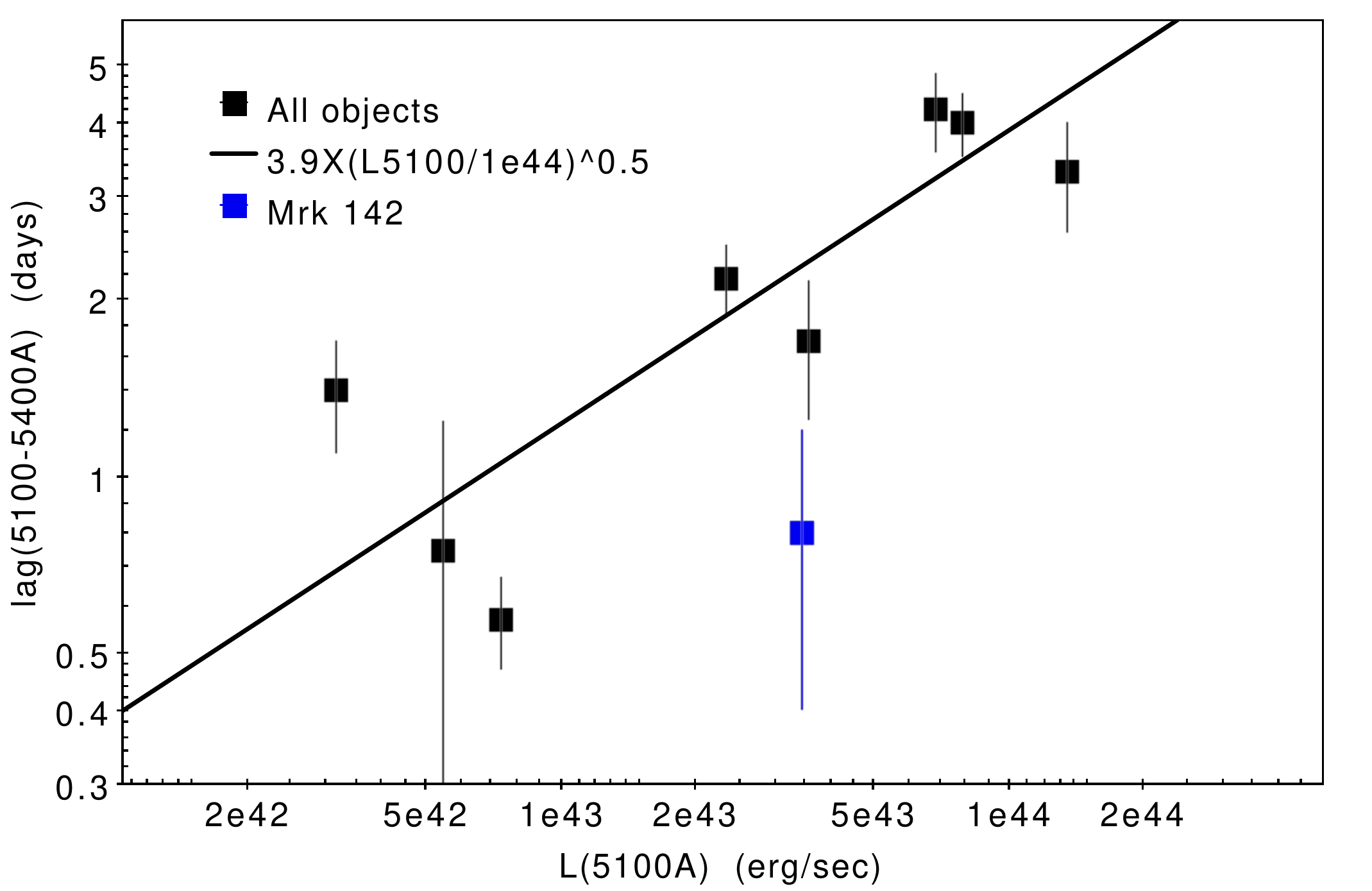}
\caption{$\tau_{5100,cont}$ vs. \Lop\ for the entire sample.
 Mrk~142 is marked in blue. 
}
\label{lag_vs_L5100}

\end{figure}

 I also used the  observed $\tau_{5100,cont}$, and the measured uncertainties, to test for a  correlation with \Lop. The results are sown in  
Fig.~\ref{lag_vs_L5100}. The line shown in the diagram is 
eqn.~\ref{rm_5100} with $c_f=0.2$ and no other adjustments. The predicted \Lop$^{1/2}$ slope gives
a very good agreement with the observations with only one source, Mrk~142,  that deviates from the curve by more than a factor 2.
Continuum RM by \cite{Cackett2020} noted an excess lag in the $U$ band of this source relative to a $\lambda^{4/3}$ line, and also relative to 
a $\lambda^{2}$ line expected in slim accretion disks. However, they did not take into account
the clear two-humps lag-spectrum which is very similar to the other group-A sources listed in table~1. The source
is one of the most extreme 
 super-Eddington AGN. Such sources are known to show large deviations from the canonical $\tau(\hb)$-\Lop\ relationship \citep[][and references 
 therein]{Du2015}. It seems that a similar deviation also exists for  $\tau_{5100,cont}$ vs. \Lop. 
 
 I have also tested
 the correlation of $\tau_{7500,cont}$ with \Lop. The model prediction is 
$\tau_{7500,cont}/\tau_{5100,cont} \approx 1.8$ and a slope 0.5 line shifted up by this amount gives a good fit to the data.
The exception, again, is Mkn~142 which lies well below
the predicted correlation. Unfortunately, $\tau_{7500,cont}$ observations are available for only 7 sources which makes this result not very meaningful.

\begin{figure} \centering
        \includegraphics[width=0.95\linewidth]{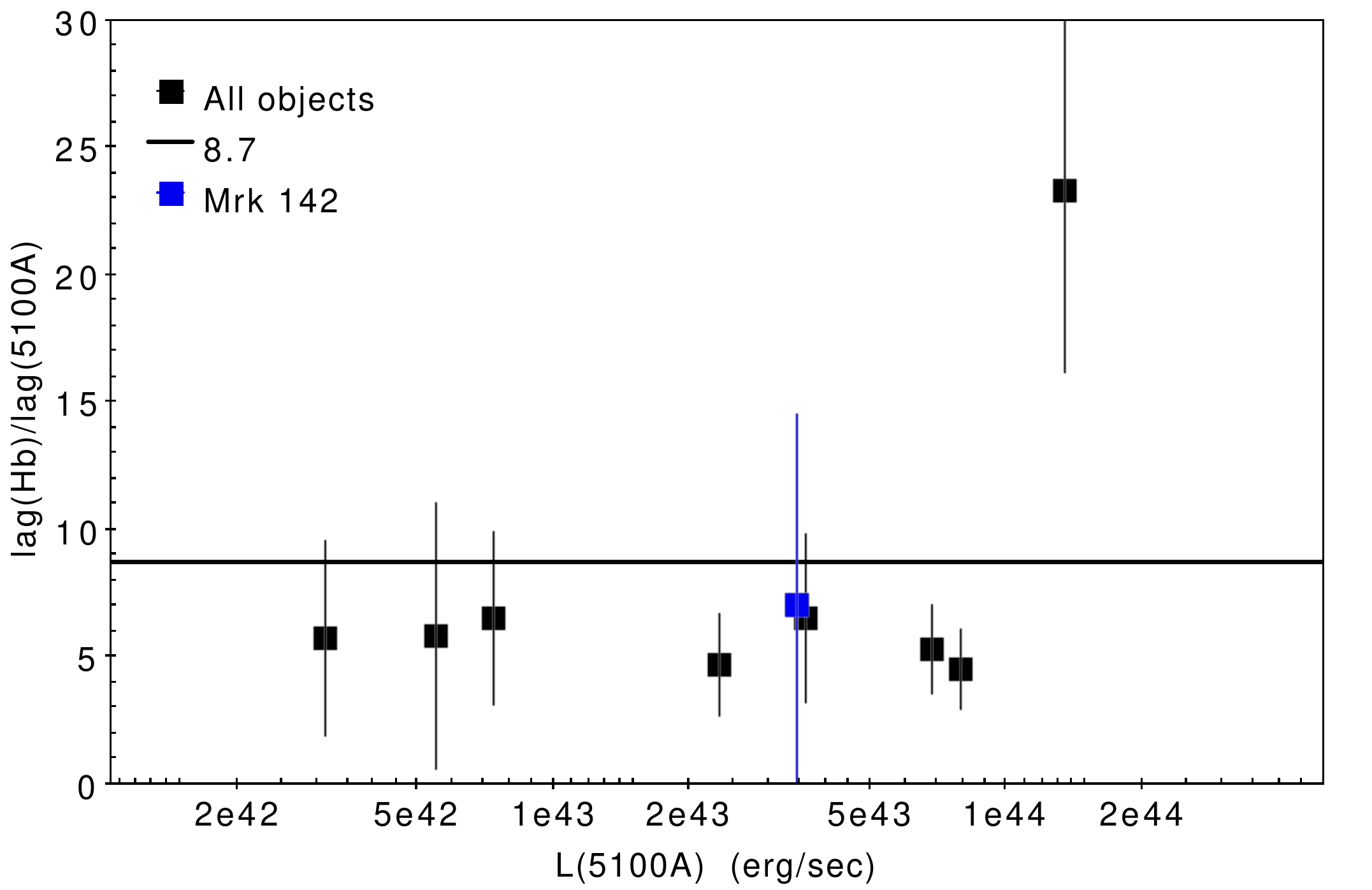}
\caption{ $\tau(\hb)/\tau_{5100,cont}$ vs. \Lop\ for all sources. Mrk~142 is marked in blue. The source close 
to the top right corner is Mrk~509 (see text for details). 
}
\label{hb_lag_over_continuum_lag}

\end{figure}

The new results can further be tested by computing $\tau(\hb)/\tau_{5100,cont}$ which is predicted to be approximately 
$ 2(1+1.5 c_f)/1.5 c_f$
with no dependence on \Lop.
 For $c_f=0.2$, this ratio is about 8.7. Fig.~\ref{hb_lag_over_continuum_lag} shows this ratio 
for all objects. Except for Mrk~509, the ratio  
is basically independent of source luminosity and is within a factor $~1.5$ of the this prediction. 
This time, Mrk~142 is not an exception confirming the smaller BLR in this source. Mrk~509 is a factor 2-3 above the line of constant ratio.
This object is well known for being above the \cite{Bentz2013} relationship by a factor $\sim2$. Interestingly,
this is not the case for its $\tau_{5100,cont}$. Since the \hb\ observations by \cite{Kaspi2000} were taken more than 15 years before the continuum lag measurements of \cite{Edelson2019}, it
is possible that the \hb\ emission region was much larger in the past, as is common for many AGN with similar luminosities. 

Finally, there are additional, mostly minor complications that must be taken into account when using such a simplified approach, e.g. the dependency on BH mass (see Fig.~\ref{m7_m8} below).

\subsection{Hot dust}

Time dependent continuum due to reverberation of the dust emission from the torus, can also contribute to the observed continuum lags, especially at long
wavelengths.  
This has been discussed, extensively, mostly with regards to the variations of the K-band flux \citep[e.g.][]{Koshida2014}.
Here I address only the contribution of this flux to the observed lags in the 1000-10000\AA\ wavelength range.

I have used the observations of \cite{Mor2012} and \cite{Lani2017} to scale the mean NIR luminosity relative to \Lop.
The mean observed ratio is $\lambda L_{NIR}$/\Lop$\approx 1.25$, where
 $\lambda L_{\lambda}$ is roughly constant over the range of 2--10$~\mu m$. The scatter around the mean is of order 0.2 dex.
 I then used a normalized $T=1700$~K greybody, about the highest dust temperature that contributes significantly to the K-band emission, 
 to estimate the contribution of such dust to the observed lag. This contribution is given by:
 \begin{equation}
  \tau_{\lambda,dust} \approx 100  L_{5100,44}^{1/2} \frac{L_{dust} }{ L_{incident}+L_{dust} } \,\, {\rm days},
 \end{equation}
 where the constant  $100  L_{5100,44}^{1/2}$ is obtained from the mean observed ratio of $\tau_{dust}/\tau(\hb) \sim 3$
 \citep[e.g.][]{gravity2020}. Scattering by dust grains in the torus is not included in the present calculations. 
 
The ratio $L_{dust}/(L_{incident}+L_{dust})$ is shown with a dashed line 
 in Fig.~\ref{diff_over_total}. The estimate is independent of the covering factor of the torus which is already included in the normalization of the plot. 
 However, $L_{dust}$ may be overestimated since a considerable fraction of the K-band emission is contributed by lower temperature dust.
 
 The above simplistic approach does not include other effects like 
  a special torus geometry, or a combination of a torus and a dusty polar wind \citep{Honig2013}. It shows that hot dust contribution to 
  the measured lag can be important at long enough wavelengths, although its predicted contributions seem to results in lags that are too long compared with the observations at $\lambda \ge 7500$\AA.

\section{Discussion}
\label{discussion}

The lag-spectra of 6 out of the 9 objects studied here clearly show broad features resembling
commonly observed spectral features in type-I AGN. While indications for the shorter wavelength feature were noted in earlier works,  and were connected to Balmer continuum
emission from the BLR, the second feature was unnoticed so far. Moreover, the overall wavelength dependence was masked by large observational uncertainties as well as attempts 
to fit lag-spectra with various illuminated accretion disk models that predict smooth dependence on wavelength. In this work I show that the two
broad features are naturally explained by combining the lag of the BLR gas to ionizing continuum variations and are not necessarily the results of the disk illumination. 
I also show a new correlation between $\tau_{5100,cont}$ and \Lop\ which looks like a scaled down version of the known $\tau(\hb)-$\Lop\ correlation and has the 
same dependence on \Lop$^{1/2}$. Finally, there is some evidence that lags due to hot dust emission may contribute too at long enough wavelengths.

The model presented here includes several simplifications.  The most important ones are the assumptions that a single emission region
can represent the entire BLR, the neglect to include, accurately, the contributions of broad emission lines to several of the bands, and the possibility that the irradiated disk contribution is larger than assumed in eqn.~\ref{disk_standard}.

As for the observations, these are limited by the small sample size (see comments in \S2), and the fact that there are only
6 sources with lag measurements that are accurate enough to perform such tests. In addition, variability-related issues suggest that the uncertainties attached to the lags are too optimistic mostly because \Lop, $\tau$(\hb) and $\tau_{5100,cont}$ are not
always measured at the same epoch. 

\subsection{Comparison with detailed BLR models}\label{section:comparison_detailed_models}

Diffuse emission lags in the 1000-10000\AA\ wavelength range combine contributions from small distances,  
where lines like \heii\ are formed (small contribution to the lags), intermediate distances where
most of the bound-free continua are produced (the largest contribution to $L_{diff}$), 
and large distances, where the Balmer and FeII lines are produced (important contribution to several of the bands, especially those covering the 2000-4000\AA\ bump). Some contributions depend more on the distribution of neutral and ionized columns as a function of distance and covering factor. An important example is Rayleigh scattering \citep{Korista1998} which is peaked close to the \La\ line center and is sensitive to the column density of
neutral hydrogen. 
Detailed computations of the dependence of several of these features on the distance from the central BH are provided in the RPC model of \cite{Netzer2020}. 

Alternative BLR models, e.g. the constant density and constant ionization parameter models of \cite{Lawther2018}, and the LOC model of \cite{Korista2019}, 
 have been proposed and discussed. Both types of models are very different from the RPC model used here. The wavelength dependencies due to  DC emission in some of them are similar to the ones shown here. In others, they are very different. As explained in \cite{Netzer2020}, some calculated time-lags in these models, especially for the Balmer lines, are much longer than assumed here. This is the results of the  different physical assumptions about the cloud properties and the much larger dust-free region assumed in several cases. 
Predicted continuum RM of such models are still to be compared with observations of other group-A sources.

\subsection{Lag-spectra and accretion disk SED}

\begin{figure} \centering
        \includegraphics[width=0.95\linewidth]{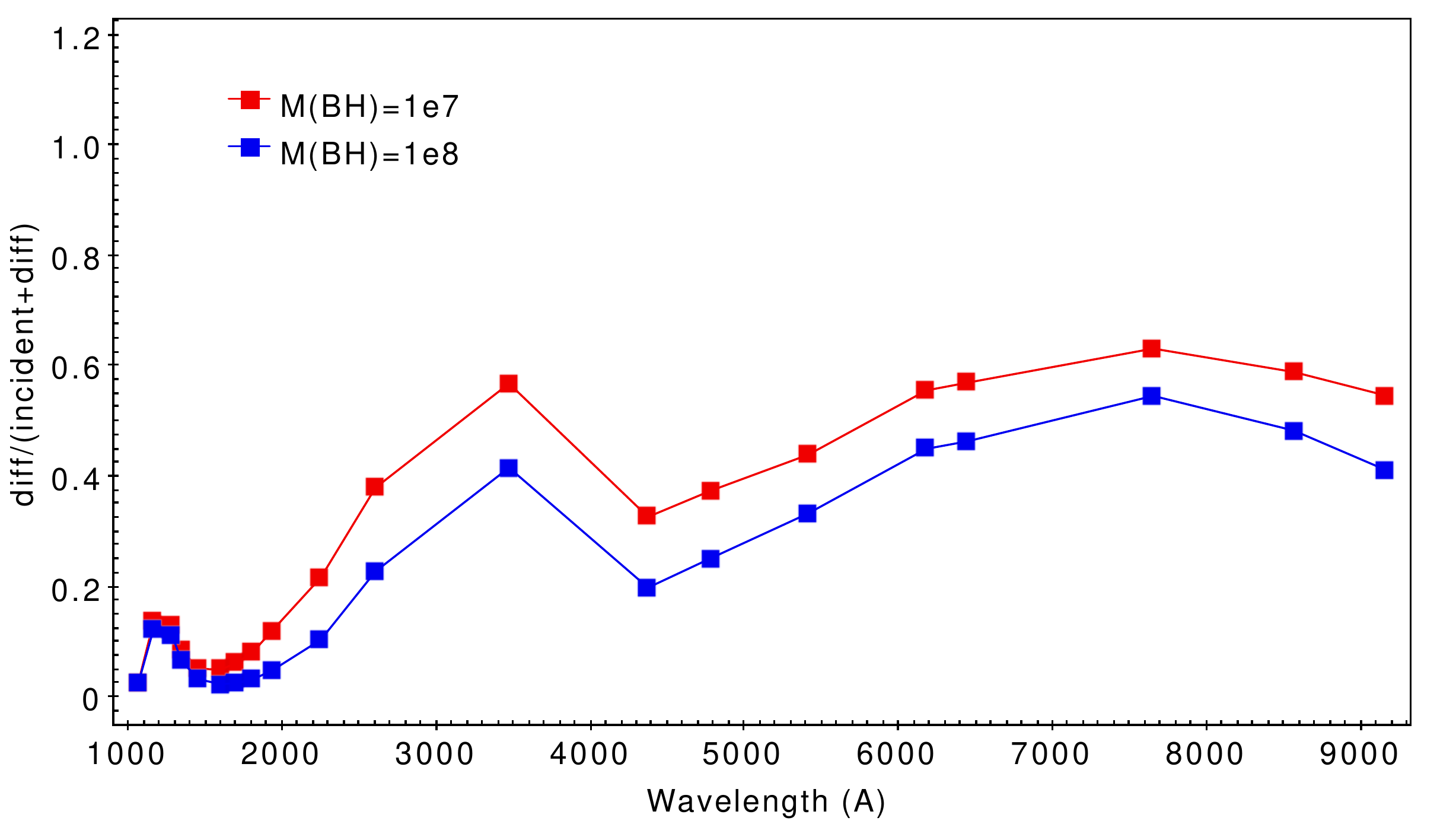}
\caption{A comparison of normalized diffuse emission lags (eqn.~\ref{rm_balmer}) between accretion disks around $10^7$ and $10^8$ 
\Msun\ BHs, as marked. The curve for the larger BH is identical to the one shown in  Fig.~\ref{diff_over_total}.
The normalized $r_{DC}$ and the BLR covering factors are the same in both cases. The disk around
the smaller BH induces longer lags  at all wavelengths above 1000\AA\ because of its steeper $\lambda > 1000$\AA\ continuum. The strong bump at 1100-1400\AA\ is mostly due to Rayleigh scattering.
}
\label{m7_m8}

\end{figure}

The disk SED which depends on BH mass, spin and accretion rate, can make a significant difference to the calculated lags. 
The SED used here assumes
an accretion disk around a BH with $a=0.7$, \MBH=$10^8$\,\Msun\ and \Ledd=0.1. Most objects in the sample have smaller BHs and probably a range of spins.
Their SEDs are, therefore, typical of higher temperature disks. To illustrate the expected changes 
I show in Fig.~\ref{m7_m8} calculations based on a model SED 
of a disk around  $10^7$\,\Msun\ BH with the same spin, covering factor, \Ledd\ and ionizing luminosity normalized to $L_{5100,44}^{1/2}$.
The lower mass results in a higher temperature and a steeper 1000-10000\AA\ continuum. The shape of the two lag-spectra are similar but the lags in the smaller BH case are longer by a wavelength dependent factor. At the longest wavelength band, this factor is about 1.5.

Fig.~\ref{m7_m8} also shows the extension of $L_{diff}/L_{incident}$  to shorter wavelengths which was not included in previous diagrams. In both cases, $\tau_{2000,cont}/\tau_{1500,cont} \approx 2$. This is similar to the observed ratio in sources where both \swift\ and \hst\ measurements are available. 
The comparison with observed lag spectra at wavelengths below $\sim 1400$\AA\ is dominated by a strong Rayleigh scattering feature. As explained in \S~\ref{section:diffuse_BLR_emission}, the treatment of this feature is complicated due to the large uncertainty on the neutral hydrogen column and the lack of a well define zero lag wavelength.

For the SED used here, and $c_f=0.2$, the fractional additional flux to the disk radiation 
 ranges from about 30\% at around 5000\AA, to about 80\% at around 7500\AA. Such contributions are not taken into account when measuring 
  the shape of the intrinsic (disk) continuum using \HST\ observations such as the ones
reported by \cite{Bentz2013} and \cite{Kara2021}. 
This can result in erroneous estimates of the incident continuum slope at wavelengths longward
of the Lyman edge. For example, if the measured
slope between $\lambda=1400$\AA\ and a certain longer wavelength indicates $L_{\nu} \propto \nu^{-0.5}$, allowing for the DC emission
would result in an intrinsic slope of $\nu^{-0.3}$ or $\nu^{-0.12}$ for 
$\lambda=5000$\AA\ and $\lambda=7500$\AA, respectively. Such slopes are in much better agreement with the typical slopes of AGN accretion disks.

The fraction of the diffuse line and continuum flux depends on the inclination angle of the disk (assumed here to be 45 degrees) since this emission is roughly isotropic. This by itself introduces a scatter of about a factor  0.7 between objects. 

There are clear indications that in some of the sources discussed in this paper (NGC~5548, Mrk~817) there are significant
contributions from powerful disk winds to the observed diffuse emission \citep[see][]{Dehghanian2020}.This emission,
and its time dependent properties, are not fully understood and are not included in the present calculations.

\subsection{FeII and high order Balmer and Paschen lines}

Strong blends of FeII emission lines are a common signature of many type-I AGN \citep[see][and references therein]{Netzer2013}. The strongest lines are those emitted over the 2000-3500\AA\ band 
and the somewhat weaker blends on both sides of the \hb\ line \citep[e.g.][]{Wills1985}.
The single component model presented here cannot reproduce the intensities of those lines whose mean emissivity radius is somewhat larger than the one for the Balmer lines. A more detailed model would show them as a strong extension of the Balmer continuum towards shorter
wavelengths with a lag which is similar to that of the Balmer continuum because of several different effects:
The combined luminosity is similar to the Balmer continuum luminosity, the line formation zone is about twice as far, and the line response to the changing continuum is weak and sub-linear. 
Detailed calculations of these 
lags are beyond the scope of the present paper.

All lag spectra shown here appear steeper than observed  
longward of the Balmer and Paschen jumps. 
These are the bands where high order Balmer and Paschen lines would normally merge, smoothly, into the respective continua. Such lines, that are 
unresolved in most cases, are  present in typical BLR spectra \citep[see many detailed 
examples in][]{Wills1985,Mejia2016}.
However, their  calculated luminosities by {\it Cloudy} fall short of the observed features by large factors.
This is related to the unexplained weakness of the calculated broad \ha\ and \hb\ lines discussed in \cite{Netzer2020}. It is likely 
the result of the simplified radiation transfer method used in {\it Cloudy}  and similar photoionization codes that fails to reproduce
the hydrogen line spectrum for conditions combining high density with high Balmer line optical depths. 

Examples showing the weakening of the Balmer lines as a function of density and optical depth are presented in appendix B. 
I also show that the calculated luminosity of the Balmer and Paschen continua are much less affected under such conditions. Thus, the calculated EW(\Hb) and other Balmer lines for $c_f=0.2$ are much smaller than observed but L(Balmer continuum)/\Lop\ is in good agreement with the observations. 
The expected time-lags for the high order lines are similar to the \hb\ lag which is about twice as long as the Balmer and Paschen continua. This will be reflected in the lag spectra over the wavelengths in question.  

Turbulent velocity inside the clouds \citep[e.g.][]{Bottorff2002a} works in exactly the opposite way. It decreases line optical depth, increases line intensities, and
completely change the structure of the clouds. 
As shown in appendix C, such an assumption is inconsistent with the RPC model considered in this paper.

\subsection{Contributions from illuminated disks}

The newly found lag-luminosity relationship is not necessarily in contradiction with the predicted dependence of $\tau_{\lambda,irr}$ on $[M \dot{M}]^{1/3}$ (eqn.~\ref{disk_standard}). This was tested for the 9 objects by using the mass, 
accretion rate and $f_{irr}$ listed in table~1. A weak correlation was found for 7 of the sources while the other two are located more than 0.3 dex below the line. However,  the lags predicted by this expression
 are 
too short compared with the observations.

The relative intensity of the irradiated to local disk flux
is about $f_{irr}/3$ and is, therefore, barely detectable in most objects and certainly below the noise level
in objects like Mrk~142 (see table~1). Increasing $\eta$ could increase the ratio but will, at the same time, shorten the lags.
Most importantly, $f_{irr}$ is predicted to be wavelength
independent which is in clear contrast to the observations that show much larger amplitude variations at shorter wavelengths. 
The longest observed lags can perhaps be explained by the \cite{Kammoun2021a} model but not the fractional flux variations. Combining with the previous discussion, one cannot reject the possibility that  X-ray irradiation can contribute, slightly, to the observed
lag-spectra that are probably dominated by diffuse, time variable BLR emission.

\subsection{Consequences to AGN accretion disks and BLR structure}

The main finding of this paper is related to the variations of the diffuse emission from the BLR which seems to explain most, if not all of the so-called IBRM. 
This does not exclude some contributions from disk illumination which implies disk size which is 2-4 times larger than the canonical \cite{Shakura1973} accretion disk.  
Independent evidence from microlensing also suggest disks that are larger than the standard optically thick geometrically thin accretion disks \citep[e.g.][]{Blackburne2011}.
Thus, there seems to be a tension between RM as explained here and microlensing observations.

Mapping the region producing the high ionization emission lines has been a problem for sometime since it seems to indicate sizes that are of the same order as the size 
of the central disk \citep[e.g. the  \heii$\lambda 4686$ emission region, see][]{Bentz2021}. A related issue is the very different lags measured for the
two \heii\ lines at 1640 and 4686\AA\ \citep{Bottorff2002b}. These puzzling issues are partly solved when taking into
account the lag of the 5100\AA\ continuum. Adding the typical numbers deduced in this paper
for the lag of the 5100\AA\ continuum (about 4 ld for a source with \Lop=$10^{44}$~\ergs), one finds high ionization emission regions
which far exceeds the size of the disk and similar lags for the two \heii\ lines.
This is directly confirmed
by the \cite{Pei2017} observations of NGC\,5548.

\section{conclusions}

This paper discusses the signature of diffuse BLR gas in the lag spectra of 9 AGN.  
The main results can be summarized as followed.
\begin{enumerate}
\item
	The lag spectra of 6 AGN with high cadence, high quality observations clearly show two features
		that resemble well known spectral feature in type-I AGN.  There are maximum time lags at around 3500 and 7500\AA\
		and a local minimum at around 5000-5500\AA.
\item
	Radiation pressure confined BLR clouds, with $c_f \sim 0.2$, result in lag-spectra which are in good agreement, in shape and in magnitude, with those observed in 5 sources. They also correctly predict the 5000-5500\AA\ lag for the three sources with lower quality observations.  The lags are scaled with the covering factor and with \Lopna\ 
	with a possible minor dependence on BH mass.
\item
	 I present a new $\tau_{5100,cont}$-\Lop\ relationship which is similar in slope
		to the well know $\tau(\hb)$-\Lop\ relationship and is scaled down by about a factor 6. This gives further support to the suggestion that most observed continuum lags are due to
		the variable diffuse BLR emission. 
\item
	Comparison of the calculated intensities of the high order Balmer and Paschen lines with the observations suggests that
		the model under-predicts their intensity by a large factor. This is similar to the general problem of the hydrogen Balmer lines in most BLR models.
\end{enumerate}

\section{Data availability}
The data used in this paper were obtained from two sources: tables in published articles and 
The AGN BH Mass Database ~http://www.astro.gsu.edu/AGNmass/.
This is listed, object by object, in table~1 where a full list of references
 is provided.

The photoionization calculations were done with \Cloudy\  which is an open source code detailed in \cite{Ferland2017}.

\section{acknowledgements}
 I thank the referee, Kirk Korista, for many useful comments and suggestions that improved the quality of this paper. I also thank Doron Chelouche, Rick Edelson, Gary Ferland and Elias Kammoun for their useful comments.

\bibliographystyle{mnras}

\newpage

\appendix

\section{Diffuse BLR spectra at different distances}

\begin{figure} \centering
        \includegraphics[width=0.95\linewidth]{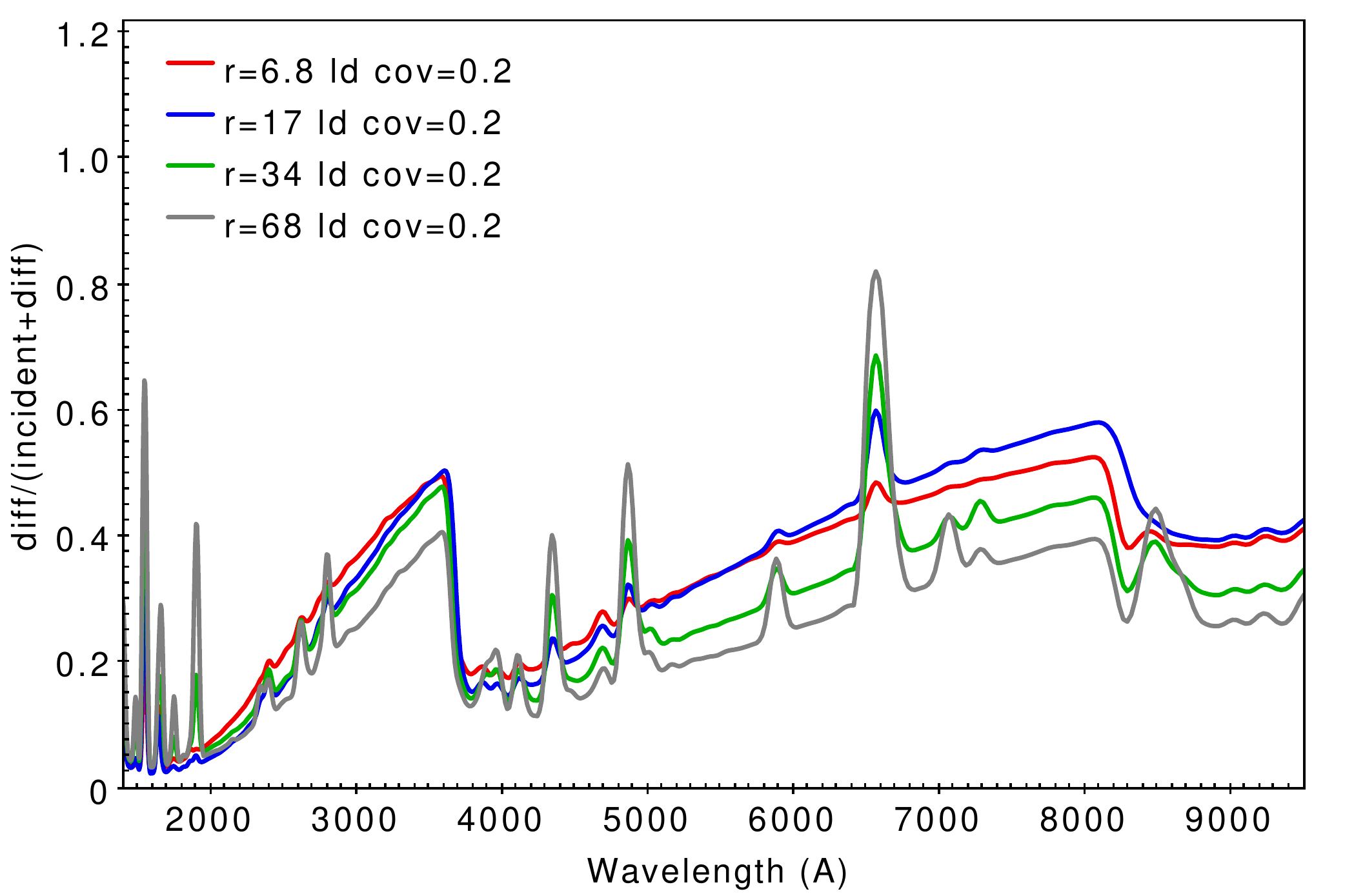}
\caption{$L_{diff}/L_{total}$ at various distances in the standard \Lopn=1 RPC model. Note the relatively small variations with distance. The model at 17 ld from the center is the one used to calculate all lag-spectra in this work. 
}
\label{appendix_A1}

\end{figure}

\begin{figure} \centering
        \includegraphics[width=0.95\linewidth]{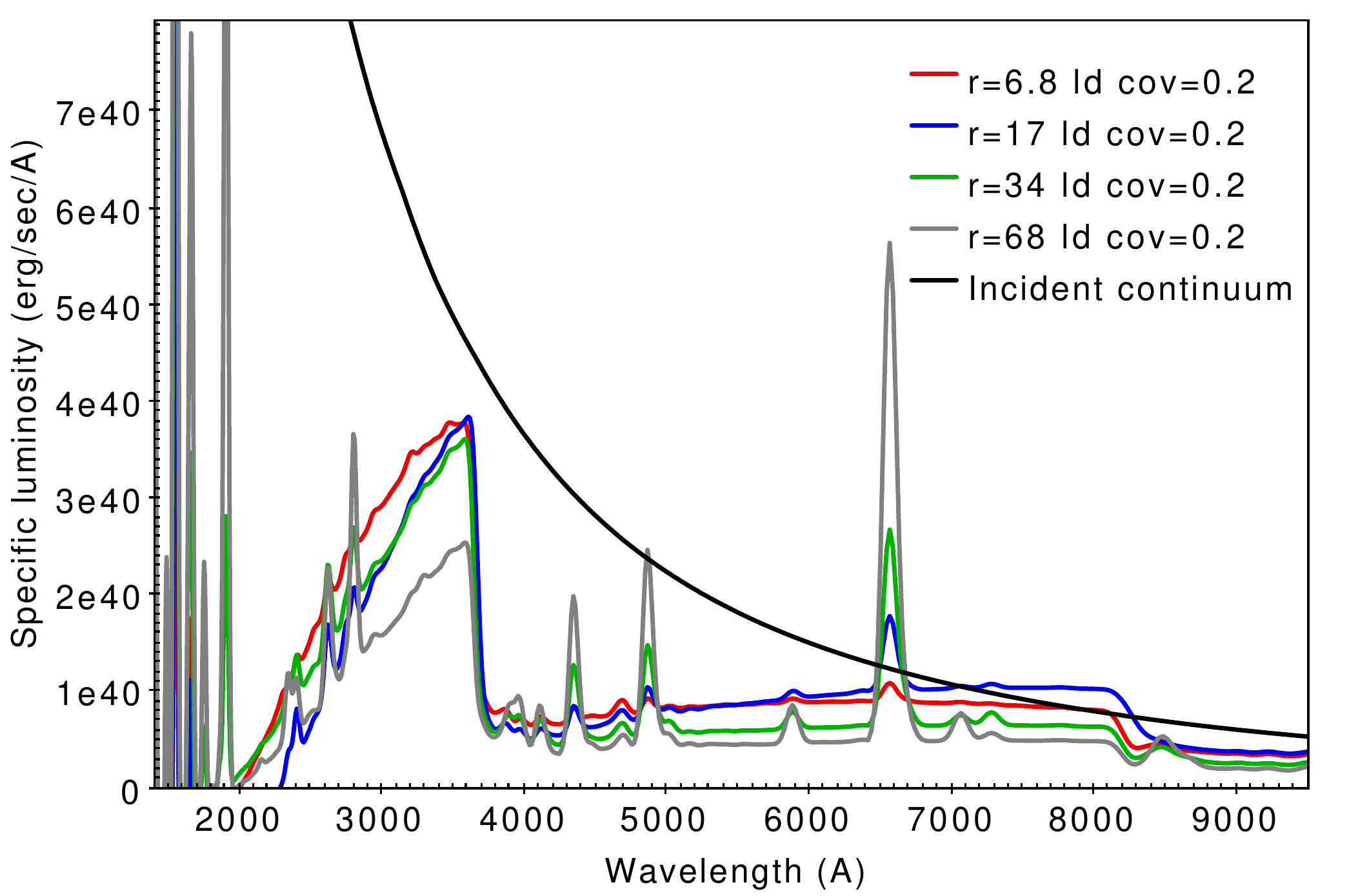}
\caption{$L_{\lambda}$ vs. $\lambda$ at various distances in the standard \Lopn=1 RPC model.  
}
\label{appendix_A2}

\end{figure}

Diffuse spectra at different distances in the standard RPC model are shown in two ways:  lag-spectra as used in this work (Fig.~\ref{appendix_A1}), and standard spectra, $L_{\lambda}$ vs. $\lambda$  (Fig.~\ref{appendix_A2}).  

\section{High order Balmer lines and the Balmer continuum}

\begin{figure} \centering
        \includegraphics[width=0.95\linewidth]{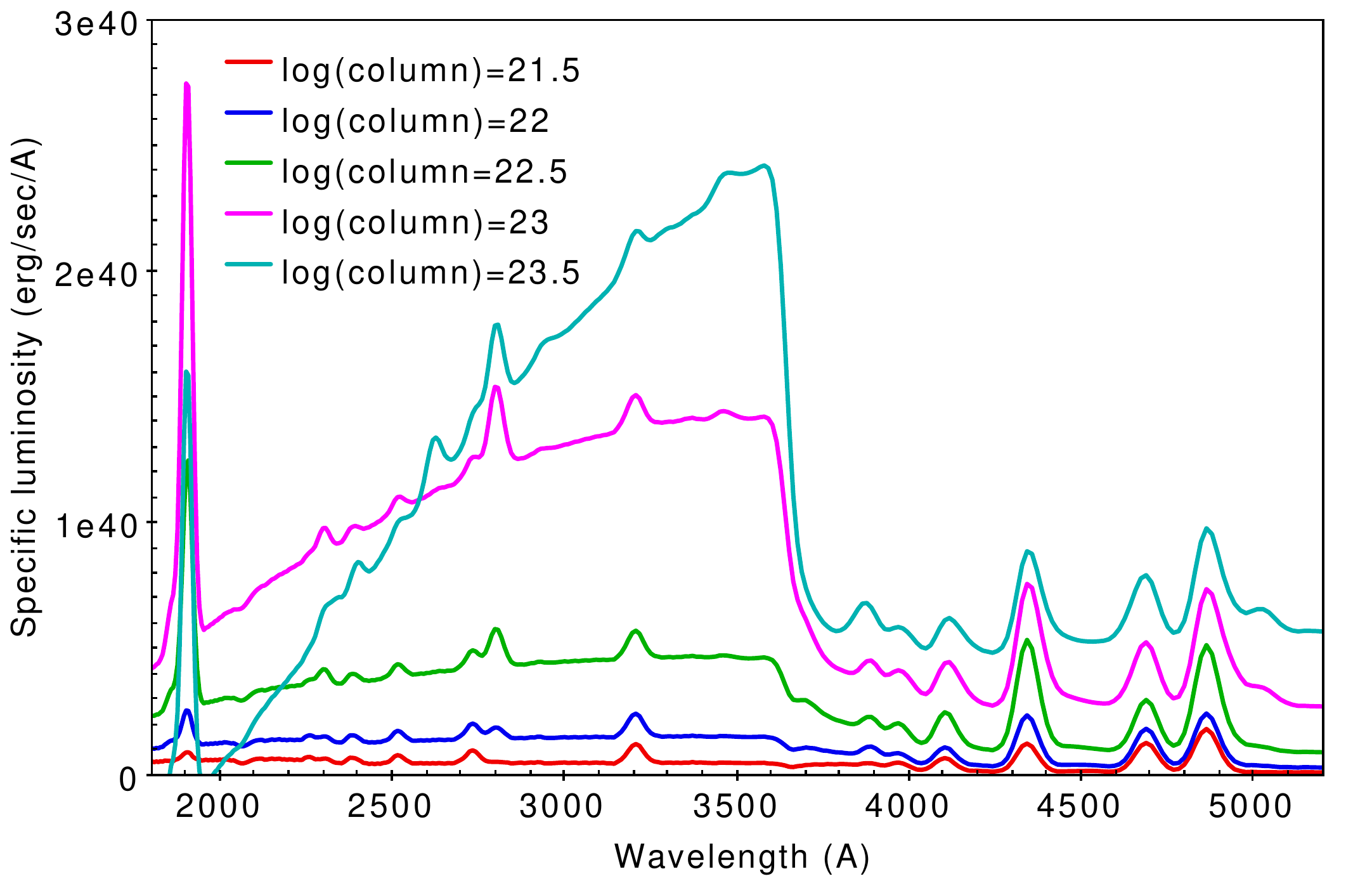}
\caption{Continuum subtracted 1500-5000\AA\ $c_f=1$ spectra of constant density clouds with various column densities, as marked,
located at the standard distance of 17\Lopna\ ld. Note the weakening of the Balmer lines relative to the Balmer continuum
with increasing column density. 
}
\label{appendix_2}

\end{figure}

Calculated luminosities of Balmer lines, and the Balmer continuum, are shown in Fig.~\ref{appendix_2}.
They assume constant density clouds, with $n_H=10^{11}~ cm^{-3}$, and various column densities, 
as marked. The clouds are 
situated at the standard distance of 17\Lopn$^{1/2}$ ld from the nominal AD1 SED. Constant density was chosen since RPC clouds 
are gravitationally bounded and hence must have very large column density, of order $10^{23.5} ~cm^{-2}$. 
The mean density of RPC clouds at this distance is somewhat higher than $10^{11}~ cm^{-3}$ but the differences in the lag-spectrum are small.
The diagram shows the decrease of the Blamer line intensities, relative to the Balmer continuum,
with increasing column density.

\begin{figure} \centering
        \includegraphics[width=0.95\linewidth]{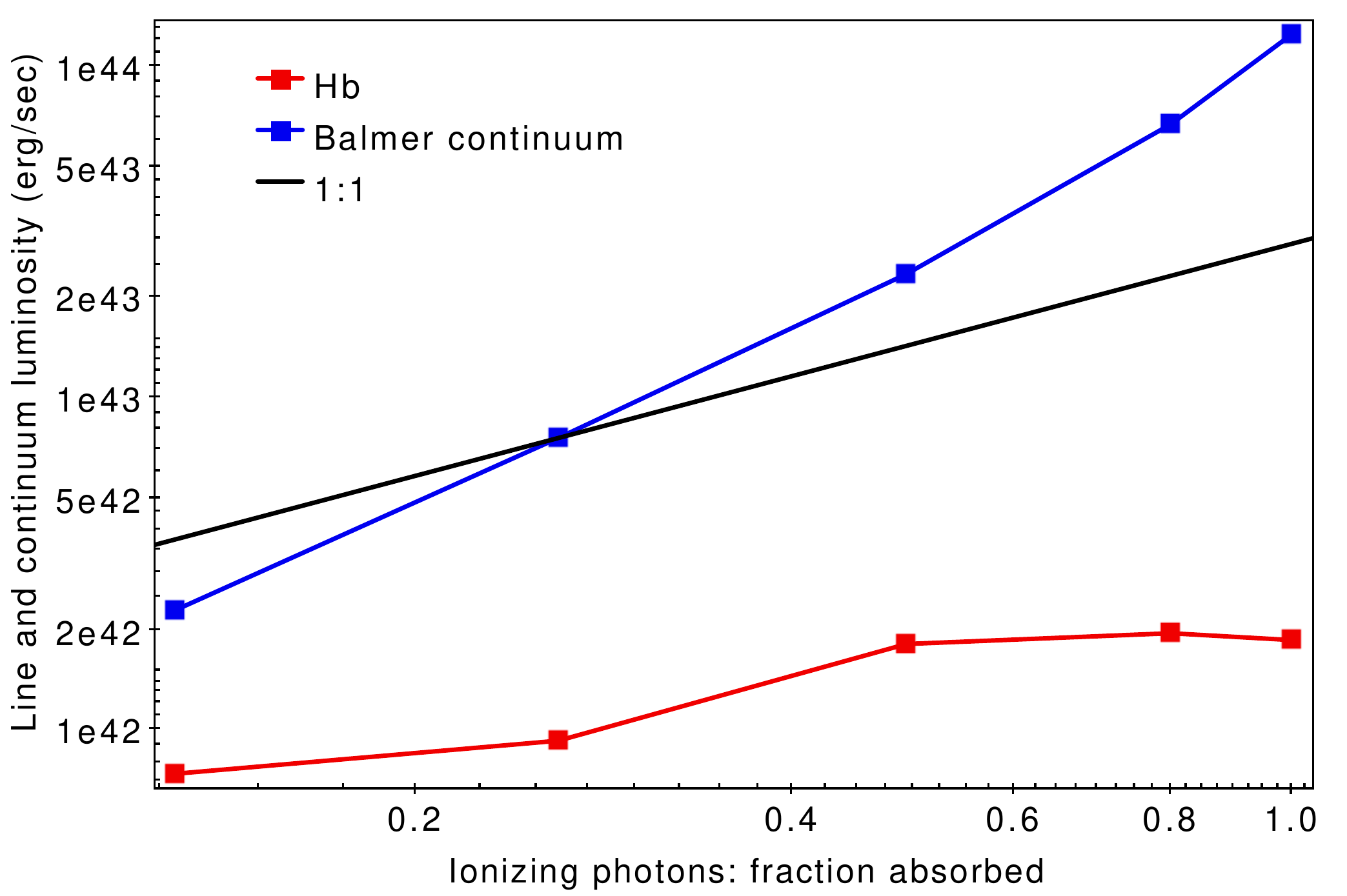}
\caption{ $L(\hb)$ and $L$(Balmer continuum) vs. the fraction of ionizing photons absorbed by 
a cloud situated  17\Lopna\ ld from the standard SED considered in this paper. Each point represents a complete 
 $c_f=1$ model. The column density increases from left to right. Note that the saturation of $L(\hb)$ starts where the fraction 
 absorbed by the gas is as small as 50\% compared with the super-linear increase of $L$(Balmer continuum) all the way to 100\% absorption. 
}
\label{appendix_3}

\end{figure}

Fig.~\ref{appendix_3} shows the formation efficiency of Balmer line and the Balmer continuum, versus the fraction of ionizing
photons absorbed by the clouds.  $L(\hb)$ increases with the absorbed fraction up
to about 50\% where it saturates, indicating line destruction due to a combination of high gas density and large optical 
depth. $L$(Balmer continuum) increases with the number of absorbed photons all the way to 100\%
The maximum \hb\ equivalent width for $c_f=1$ is about 79\AA. The high order Balmer lines (not shown here) behave in a way similar to \hb.

\begin{table}
\centering
	\caption{
		Comparison of the standard single location model of this work with a full-BLR model.
	}
\begin{tabular}{lccc}
Property$^a$         & Present Model & \cite{Netzer2020}$^b$ & Observed$^c$  \\
\hline
L(\hb)/\Lop\         & 0.003          & 0.0034                      & 0.02     \\            
$L$(Balmer)/\Lop$^d$   & 0.23           & 0.29                        & 0.22 \\                 
\hline
\end{tabular}
\begin{tablenotes}
\item
$^a$ All properties are normalized to $c_f=0.2$
$^b$ AD1-SED p=2.4 model in Table 3 normalized to $c_f=0.2$.
$^c$ Assuming Balmer continuum emission contributes 50\% of the luminosity of the 
2000-4000\AA\ bump.
\item
$^d$ Main references: \cite{Wills1985}, \cite{Netzer1985}, \cite{Mejia2016}.
\end{tablenotes}
\label{tab:appendix}
\end{table}

\section{Turbulence}
Turbulent motion inside the BLR clouds increase the internal pressure, decrease all line optical depths, and reduce the destruction of line photons \citep{Bottorff2002a}.
This increases the line luminosity and changes the internal energy budget. 

\begin{figure} \centering
        \includegraphics[width=0.95\linewidth]{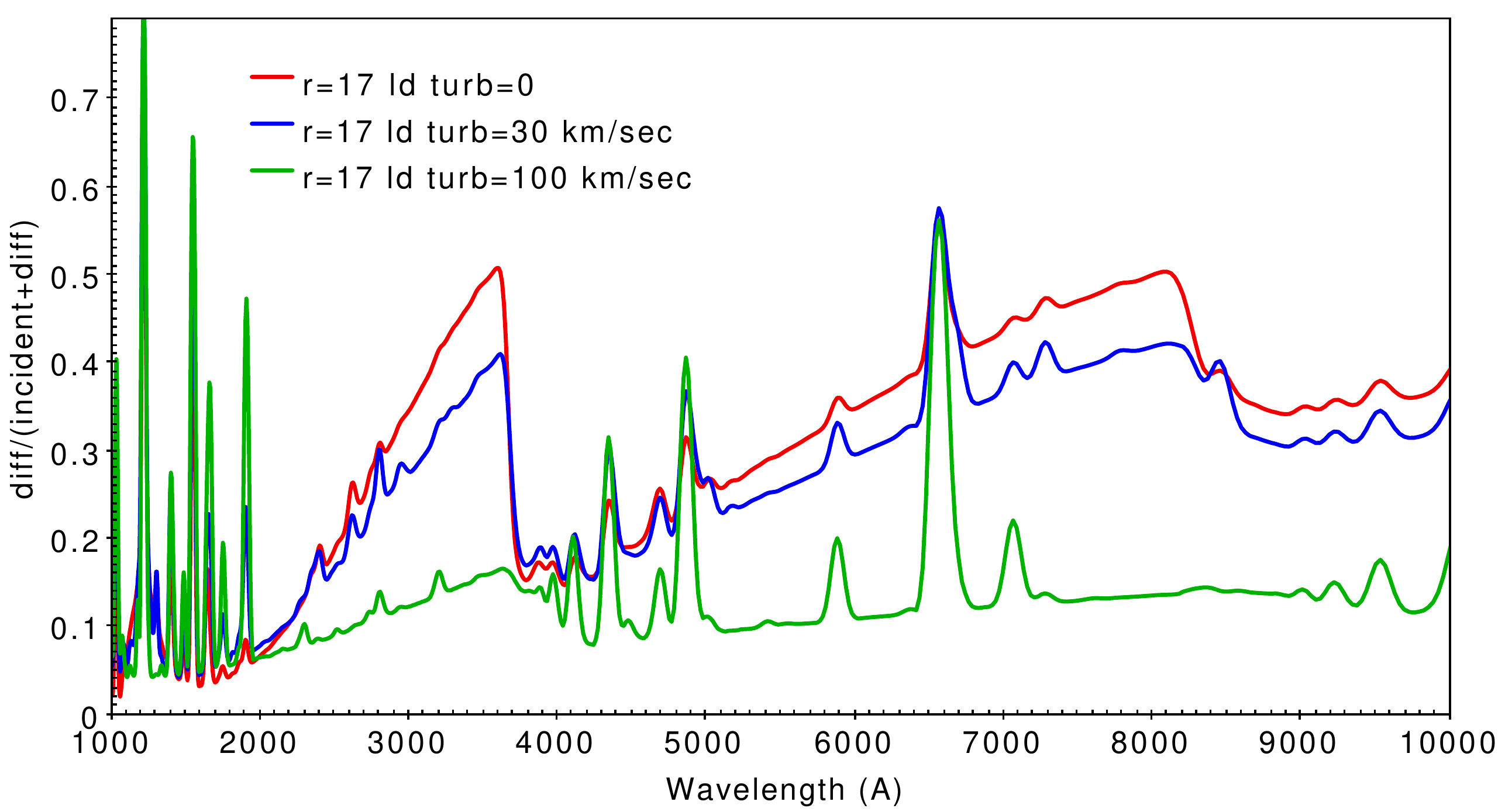}
	\caption{$L_{diff}/L_{total}$ for clouds with various internal turbulent velocities, as marked (the red curve is the standard model discussed in this paper) and gas density of $\sim 10^{10}$~\cc\ at the illuminated face.
	The luminosity of all Balmer lines increase with increasing turbulent velocity and the pressure inside the cloud deviates more and more
	from the pressure in the canonical RPC model due to turbulent pressure, $p_{turb}$. 
	For $v_{turb}$=30 \kms,  
	 $p_{turb}$ is comparable to $p_{gas}$ close to the ionization front.
	For $v_{turb}$=100 \kms, $p_{turb}/p_{gas} \approx 20$(!).
}
\label{appendix_4}

\end{figure}

Fig.~\ref{appendix_4} shows two examples of constant pressure clouds with gas density at the illuminated face of $\sim 10^{10}$~\cc. In one case $v_{turb}=30$ \kms\ and in the second case
$v_{turb}=100$ \kms. 
The emitted spectrum, especially in the second case, is, indeed very different, since the cloud
remains completely ionized even at the large column density assumed here ($10^{23.5}$~\cmii). This results in stronger emission lines and weaker diffuse continuum.
In the second case the internal gas pressure is only a small fraction of the total
pressure (see figure caption). 
Raising the gas density to $10^{11}$~\cc\ results in a large column of neutral gas at the back
of the cloud but a small change in the spectrum. Finally, 
turbulence requires an additional source of dissipative heating which is not part of the energy budget considered here.
It is thus difficult to construct well motivated, physically justified photoionization models for the entire 
BLR with large internal turbulent velocity in some or all the clouds. 
All this is inconsistent with the RPC model which is the main assumption of this paper.

\end{document}